\documentclass[twocolumn,twocolappendix,tighten]{aastex63}

\received{6 March 2023}
\revised{30 May 2023}
\accepted{27 June 2023}
\submitjournal{ApJ}

\pdfoutput=1 %for arXiv submission
\usepackage{amsmath,amstext}
\usepackage[T1]{fontenc}
\usepackage[figure,figure*]{hypcap}

\usepackage{graphicx}
\usepackage{graphics}
\usepackage{color}
\usepackage{lipsum}
\usepackage[caption=false]{subfig}
\definecolor{dartmouthgreen}{rgb}{0.05, 0.5, 0.06}
\usepackage{afterpage}
\usepackage{natbib}
\usepackage[percent]{overpic}

\def\arcsec{\hbox{$^{\prime\prime}$}}
\def\deg{\hbox{$^\circ$}}
\def\hr{\textsuperscript{h}}
\def\min{\textsuperscript{m}}
\def\sec{\textsuperscript{s}\hspace{-0.7mm}}

\def\nh{$N_{\rm H}$}
\def\nhm{\textit{N}_{\rm{H}}/\rm{cm}^{-2}}

\def\nh{$N_\mathrm{H}$}

\def\chandra{\textit{Chandra}}
\def\xmm{\textit{XMM-Newton}}
\def\wise{\textit{WISE}}
\def\nustar{\textit{NuSTAR}}

\def\sdss{\textit{SDSS}}

\graphicspath{ {Images/} }

\shorttitle{NuSTAR Observes Mid-IR Dual AGN Candidates}
\shortauthors{Pfeifle et al.}
\begin{document}
\title{\nustar{} Observations of Four Mid-IR Selected Dual AGN Candidates in Galaxy Mergers}

\correspondingauthor{Ryan W. Pfeifle}
\email{ryan.w.pfeifle@nasa.gov}

\author[0000-0001-8640-8522]{Ryan W. Pfeifle}
\altaffiliation{NASA Postdoctoral Program Fellow}
\altaffiliation{Surname pronunciation: ``\textit{Fife-Lee}''}
\affiliation{X-ray Astrophysics Laboratory, NASA Goddard Space Flight Center, Code 662, Greenbelt, MD 20771, USA}
\affiliation{Oak Ridge Associated Universities, NASA NPP Program, Oak Ridge, TN 37831, USA}
%\affiliation{Department of Physics and Astronomy, George Mason University, 4400 University Drive, MSN 3F3, Fairfax, VA 22030, USA}

\author{Kimberly Weaver}
\affiliation{X-ray Astrophysics Laboratory, NASA Goddard Space Flight Center, Code 662, Greenbelt, MD 20771, USA}

\author[0000-0003-2277-2354]{Shobita Satyapal}
\affiliation{Department of Physics and Astronomy, George Mason University, 4400 University Drive, MSN 3F3, Fairfax, VA 22030, USA}

\author[0000-0001-5231-2645]{Claudio Ricci}
\affiliation{Instituto de Estudios Astrof\'isicos, Facultad de Ingenier\'ia y Ciencias, Universidad Diego Portales, Av. Ej\'ercito Libertador 441, Santiago, Chile}
\affiliation{Kavli Institute for Astronomy and Astrophysics, Peking University, Beijing 100871, China}
\affiliation{Department of Physics and Astronomy, George Mason University, 4400 University Drive, MSN 3F3, Fairfax, VA 22030, USA}

\author[0000-0002-4902-8077]{Nathan J. Secrest}
\affiliation{U.S. Naval Observatory, 3450 Massachusetts Avenue NW, Washington, DC 20392, USA}

\author[0000-0002-8818-9009]{Mario Gliozzi}
\affiliation{Department of Physics and Astronomy, George Mason University, 4400 University Drive, MSN 3F3, Fairfax, VA 22030, USA}

\author[0000-0002-2183-1087]{Laura Blecha}
\affiliation{Department of Physics, University of Florida, P.O. Box 118440, Gainesville, FL 32611-8440, USA}

\author[0000-0003-2283-2185]{Barry Rothberg}
\affiliation{U.S. Naval Observatory, 3450 Massachusetts Avenue NW, Washington, DC 20392, USA}
\affiliation{Department of Physics and Astronomy, George Mason University, 4400 University Drive, MSN 3F3, Fairfax, VA 22030, USA}

\begin{abstract}
Mergers of galaxies are a ubiquitous phenomenon in the Universe and represent a natural consequence of the ``bottom-up'' mass accumulation and galaxy evolution cosmological paradigm. It is generally accepted that the peak of AGN accretion activity occurs at nuclear separations of $\lesssim10$\,kpc for major mergers. Here we present new \nustar{} and \xmm{} observations for a subsample of mid-IR preselected dual AGN candidates in an effort to better constrain the column densities along the line-of-sight for each system. Only one dual AGN candidate, J0841+0101, is detected as a single, unresolved source in the \xmm{} and \nustar{} imaging, while the remaining three dual AGN candidates, J0122+0100, J1221+1137, and J1306+0735, are not detected with \nustar{}; if these non-detections are due to obscuration alone, these systems are consistent with being absorbed by column densities of log($\nhm{}$) $\geq$ 24.9, 24.8, and 24.6, which are roughly consistent with previously inferred column densities in these merging systems. In the case of J0841+0101, the analysis of the 0.3-30 keV spectra reveal a line-of-sight column density of \nh{} $\gtrsim10^{24}$\,cm$^{-2}$, significantly larger than the column densities previously reported for this system and demonstrating the importance of the higher signal-to-noise \xmm{} spectra and access to the $>10$\,keV energies via \nustar{}. Though it is unclear if J0841+0101 truly hosts a dual AGN, these results are in agreement with the high obscuring columns expected in AGNs in late-stage mergers. 
\end{abstract}

\keywords{Dual AGN --- 
Galaxy Merger --- X-ray Astronomy}

\section{Introduction} \label{sec:intro}

Over the last two and a half decades we have come to understand that supermassive black holes (SMBHs) reside at the centers of most massive galaxies and that the masses of these SMBHs strongly correlate with the stellar velocity dispersions \citep[M-$\sigma$, e.g.,][]{ferrarese2000,gebhardt2000} and the luminosities \citep[M-L][]{gultekin2009} of their host spheroids. How these scaling relations between the SMBHs and their hosts are established remains an open debate, though one possible formation pathway is through the merging of galaxies, a ubiquitous phenomenon that is a key component of ``bottom-up'' mass accumulation and galaxy evolution. Gravitational tidal torques induced during a galaxy merger have been shown to drive large reservoirs of gas and dust into the galaxy nuclei \citep[e.g.,]{barnes1996}, potentially fueling both star formation \citep[e.g.,][]{barnes1991,mihos1996} and the central SMBHs \citep[e.g.,][]{hopkins2006,hopkins2008}, and thus mergers offer an efficient avenue for correlated stellar and SMBH growth. Numerous hydrodynamic simulations performed over the last decade also agree that correlated SMBH growth -- manifesting as dual active galactic nuclei (AGNs) -- are expected to occur in late-stage mergers\footnote{Note, dual AGNs can be found in earlier stage mergers \citep[e.g.,][]{guainazzi2005,koss2012,blecha2018,derosa2018}, but correlated growth is more likely in late-stage mergers.} with nuclear pair separations $<10$\,kpc and should coincide with the peak of the merger-induced SMBH growth \citep{vanwassenhove2012,capelo2015,capelo2017,blecha2018} while simultaneously being enshrouded by large columns of gas and dust \citep{capelo2017,blecha2018}. Observations agree well with these obscuration predictions; mergers have been shown to host higher fractions of obscured (and heavily obscured) AGNs compared to control samples \citep[e.g.,][]{satyapal2014,kocevski2015,ricci2017MNRAS,lanzuisi2018,koss2018,ricci2021}, and the mid-IR  AGN excess relative to optical AGNs increases as a function of decreasing nuclear pair separation, supporting the idea that obscuration increases in late-stage mergers \citep[e.g.,][]{satyapal2014}. While only a few dozen dual AGNs have thus far been confirmed in the literature, a significant fraction of these systems show direct and indirect evidence for large absorbing columns on the order of $\gtrsim10^{23}$-$10^{24}\,\rm{cm}^{-2}$ \citep[e.g.,][]{komossa2003,bianchi2008,piconcelli2010,koss2016,derosa2018,pfeifle2019a,pfeifle2019b}. These high column densities may partially explain the lack of a larger population of known dual AGNs; dual AGN candidates are most often selected using optical spectroscopic diagnostics such as double-peaked emission lines \citep[e.g.,][but with relatively little success]{wang2009,liu2010,smith2010,comerford2011,comerford2012,comerford2015,mullersanchez2015} or narrow spectroscopic emission line ratios \citep[][]{liu2011,guainazzi2021}, but large absorbing columns along the line-of-sight and/or high covering factors may bias against detecting large populations of dual AGNs in the optical band, resulting in fewer confirmed cases \citep[see discussion in, e.g.,][]{koss2012}. Hydrodynamic simulations with radiative transfer post-processing calculations performed in \citet{blecha2018} demonstrated that this obscured phase of expeditious dual SMBH growth should be traced by the mid-infrared (mid-IR) colors of the merging system, where dual AGNs should exhibit Wide-Field Infrared Survey Explorer (\wise{}) mid-IR colors consistent with $W1-W2>0.5$. Since the mid-IR colors trace the activity of the SMBHs regardless of the column density (and are therefore less sensitive to biases due to obscuration), mid-IR selection may therefore offer a more statistically complete method of searching for and identifying dual AGNs.

Motivated by these recent theoretical results, we have been studying a sample of 15 late-stage galaxy mergers pre-selected with \wise{} \citep[see][]{satyapal2017,pfeifle2019a,pfeifle2019b}. The sample selection methodology is thoroughly described in \citet{satyapal2017} and \citet{pfeifle2019a}, but we briefly describe it here: we drew our sample from the Galaxy Zoo project \citep{lintott2008},\footnote{\url{http://www.galaxyzoo.org}} from the Sloan Digital Sky Survey (SDSS) DR7 \citep{abazajian2009}, selecting only systems that were most likely to be strongly disturbed mergers based upon the Galaxy Zoo classifications \citep[weighted-merger-vote-fraction, $f_{m}>0.4$,][]{darg2010MNRAS}. Cross-matching this sample with the AllWISE \citep{https://doi.org/10.26131/irsa1} release of the \wise{} catalog\footnote{\url{http://wise2.ipac.caltech.edu/docs/release/allwise/}}, we required detections in the W1 and W2 bands with a signal to noise ratio $>5\sigma$ and we applied the dual AGN mid-IR color criterion of $W1-W2>0.5$ from \citet{blecha2018}. Visual inspection ensured the presence of two nuclei with projected separations $<10$\,kpc that were resolvable with \chandra{}. We list this sample of mid-IR selected dual AGN candidates in Table~\ref{table:sample} along with their redshifts, angular separations, and pair separations \citep[see Table~1 in][for additional details]{pfeifle2019a}. Two key questions we wished to address were: How effective is mid-IR selection in identifying dual AGNs in late-stage mergers? (2) Are dual AGNs in mid-IR selected mergers indeed heavily obscured? 

In \citet[][]{satyapal2017} and \citet[][]{pfeifle2019a}, our follow-up  \chandra{} X-ray observations revealed dual nuclear X-ray sources in 8/15 of the mid-IR selected mergers. One of the targets which met our selection criteria and was included in our sample was Mrk\,463, which actually hosts a known dual AGN system \citep{bianchi2008}. Both direct (spectral analysis) and indirect diagnostics for absorption indicated that the AGNs in these mergers were indeed heavily obscured, with column densities $\gtrsim10^{23}$-$10^{24}\,\rm{cm}^{-2}$, in agreement with predictions from simulations \citep{capelo2017,blecha2018}. However, constraints on column densities and other AGN X-ray properties in heavily obscured AGNs can be difficult to obtain with softer energy ($<10\,$keV) X-ray telescopes alone without access to harder energies and specifically harder X-ray spectral features, such as the Compton reflection hump beyond $>10$\,keV. The Nuclear Spectroscopic Telescope Array (\nustar{}) on the other hand offers access to energies beyond 10 keV \citep[half power diameter of 58'' and FWHM of 18'';][]{harrison2013}, and \nustar{} has been successful over the last decade in detecting large samples of obscured AGNs \citep[e.g.,][]{lansbury2017, ricci2017bass,marchesi2018} -- including heavily obscured AGNs in mergering systems \citep[e.g.,][]{ricci2017MNRAS,ricci2021,lansbury2017,yamada2021} -- and constraining their column densities and X-ray properties. In many cases, NuSTAR allowed for refined estimates of column densities derived from lower quality and/or softer X-ray spectra \citep[e.g.,][]{lansbury2015,marchesi2018}. Therefore, the natural next step in our analysis of these heavily obscured dual AGNs and candidates was to study their hard X-ray properties as observed with \nustar{}.

\begin{table}
\caption{WISE-Selected Dual AGN Candidate Sample}
\begin{center}
\begin{tabular}{ccccc}
\hline
\hline
\noalign{\smallskip}
Name  & z & $\Delta \theta$  & $r_{\rm{p}}$ & Class \\
\noalign{\smallskip}
(SDSS) & &  (\arcsec) & (kpc) \\
\noalign{\smallskip}
\hline
\noalign{\smallskip}
J012218.11+010025.7$^*$ & 0.05546  & 8.7 & 8.7 & DC \\
J084135.08+010156.2$^*$ & 0.11060  & 3.9 & 7.9 & DC \\
J084905.51+111447.2 & 0.07727  & 2.2$^{(a)}$ & 3.3$^{(a)}$ & T \\ 
           &                   & 4.0$^{(b)}$ & 5.8$^{(b)}$ & T\\
J085953.33+131055.3 & 0.03083  & 16.1 & 9.9 & S \\
J090547.34+374738.2 & 0.04751  & 6.2 & 5.8 & S \\
J103631.88+022144.1 & 0.05040  & 2.8 & 2.8 & S\\
J104518.03+351913.1 & 0.06758  & 7.0 & 9.0 & DC \\
J112619.42+191329.3 & 0.10299  & 2.3 & 4.5 & S\\
J114753.62+094552.0 & 0.09514  & 3.8$^{(c)}$ & 6.8$^{(c)}$ & S\\
            &                  & $2.4^{(d)}$ & $4.3^{(d)}$ & S\\ 
J115930.29+532055.7 & 0.04498  & 2.7 & 2.4 & S\\
J122104.98+113752.3$^*$ & 0.06820  & 7.1 & 9.3 & DC\\
J130125.26+291849.5 & 0.02340  & 21.8 & 10.3 & DC\\
J130653.60+073518.1$^*$ & 0.11111  & 2.0$^{(e)}$ & 4.0$^{(e)}$ & DC\\
          &                    & 3.7$^{(f)}$ & 7.4$^{(f)}$ & DC\\
J135602.89+182218.2 & 0.05060  & 4.0 & 4.0 & D\\
J235654.30-101605.3 & 0.07390  & 3.6 & 5.0 & S\\ 
\noalign{\smallskip}
\hline
\end{tabular}
\end{center}
\tablecomments{Basic information on the sample of 15 mid-IR selected dual AGN candidates from \citet{satyapal2017} and \citet{pfeifle2019a}. Column 1: SDSS target designation. Column 2: Redshifts. Columns 3$\--$4: Angular separation of the galaxy nuclei in arcseconds and kiloparsecs, respectively. Column 5: classification following our analysis in \citet{satyapal2017}, \citet{pfeifle2019a} and \citet{pfeifle2019b}; we denote duals with (D), single AGNs with (S), triple AGNs with (T), and we append (C) to the classification to indicate candidates. See \citet{pfeifle2019a} for a full version of this table. (a) Angular separation between the SW and SE X-ray sources. (b) Angular separation between the SE and N X-ray sources. (c) Angular separation between the S and NE nuclei. (d) Angular separation between the S and NW nuclei. (e) Angular separation between the NE and SW X-ray sources. (f) Angular separation between the SW and SE X-ray sources. $^*$: objects studied in this work.}
\label{table:sample}
\end{table}

Here we present new \nustar{} and \xmm{} observations of a subsample of four mid-IR selected dual AGN candidates drawn from \citet{satyapal2017} and \citet{pfeifle2019a}, obtained in an effort to constrain the column densities in these potentially heavily obscured AGNs (marked with asterisks in Table~\ref{table:sample}). This work is organized as follows: In Section 2 we outline the new \xmm{} and \nustar{} observations and detail our data processing steps. In Section 3 we describe our data analysis and in Section 4 we report the X-ray photometric and spectroscopic results. We discuss these results and their relation to the literature in Section 5 and we present our conclusions in Section 6. Throughout this work we adopt the following cosmology: $\textrm{H}_0 = 70$\,km\,s$^{-1}$\,Mpc$^{-1}$, $\Omega_M=0.3$, and $\Omega_\Lambda=0.7$.

\section{Observations and Data Reduction}
\label{sec:obsandred}

\begin{table*}
\begin{center}
\caption{New X-ray Observations Examined in this Work}
\label{table:nustarobs}
\begin{tabular}{ccccccccc} %ccccc
\hline
\hline
\noalign{\smallskip}
\noalign{\smallskip}
Name & $\alpha$ & $\delta$ & Obs. Date & ObsID & Instrument & Exp. & Net Exp. \\ %& SAA & SAA & Tent. & Source & BKG \\
(1) & (2) & (3) & (4) & (5) & (6) & (7) & (8) \\% Calc. & Mode &  & Aperture Radius & Aperture Size \\
\noalign{\smallskip}
\noalign{\smallskip}
\hline
\noalign{\smallskip}
J0122+0100 & 01\hr{}22\min{}21.\sec{}0	& +00\deg{}58\min{}39\sec{} & 26 Oct. 2018 & 60467001002 & NuSTAR FPMA & 32.4 ks & 31.9 ks \\ 
J0122+0100 & \dots	& \dots & \dots & \dots & NuSTAR FPMB & 32.4 ks & 31.7 ks \\ 
J0841+0101 & 08\hr{}41\min{}45.\sec{}3	& +01\deg{}03\min{}06\sec{} & 29 Dec. 2018 & 60401002002 & NuSTAR FPMA & 48.4 ks & 43.6 ks \\ 
J0841+0101 & \dots	& \dots & \dots & \dots & NuSTAR FPMB & 48.4 ks & 43.4 ks \\
J0841+0101 & 08\hr41\min35\sec.04 & +01\deg01\arcmin55\arcsec.2 & 10 May 2018 & 0822470101 & XMM EPIC PN & 32 ks & 28.7 ks \\
J0841+0101 & \dots & \dots & \dots & \dots & XMM EPIC MOS1 & 32 ks & 30.6 ks \\
J0841+0101 & \dots & \dots & \dots & \dots & XMM EPIC MOS2 & 32 ks & 30.6 ks \\
J1221+1137 & 12\hr{}20\min{}56.\sec{}0	& +11\deg{}36\min{}52\sec{}	& 11 May 2019 & 60467002002 & NuSTAR FPMA & 29.9 ks & 27.5 ks \\ 
J1221+1137 & \dots	& \dots	& \dots & \dots & NuSTAR FPMB & 29.9 ks & 28.8 ks \\ 
J1306+0735 & 13\hr{}06\min{}45.\sec{}2	& +07\deg{}34\min{}25\sec{} & 20 Jun. 2019 & 60467003002 & NuSTAR FPMA & 49.7 ks & 49.6 ks \\ 
J1306+0735 & \dots	& \dots & \dots & \dots & NuSTAR FPMB & 49.7 ks & 49.3 ks \\ 
\noalign{\smallskip}
\hline
\end{tabular}
\end{center}
\tablecomments{Col. $1$: Truncated merger designation. Col. $2\--3$: Coordinates of X-ray observations. Col. $4\--5$: UT date of X-ray observations and observation ID. Col. $6$: facility and camera. Col. $7\--8$: Effective exposure time after background flare filtering.}
\end{table*}

\subsection{\nustar{} Observations}
\label{sec:nustarobs}
Pointed observations of J0122+0100, J1221+1137, and J1306+0735 were obtained with \nustar{} between 26 Oct. 2018 and 20 June 2019 (Program ID 04185) while J0841+0101 was observed by NuSTAR on 10 May 2018 (Program 082247). The former three mergers were targeted as a part of a NuSTAR follow-up program\footnote{For these observations, we estimated the intrinsic AGN 2-10 keV luminosities by converting the 6\,$\mu$m luminosity (derived from WISE) to the 2-10 keV luminosity via the \citet{stern2015} relation. We used these intrinsic X-ray luminosities along with the NuSTAR responses to simulate fake spectra using MYTORUS and gauge the required exposure times needed for 400-900 counts.} for the four dual AGN candidates in our pilot program \citep{satyapal2017}; J1045+3519 was included in that proposal but unfortunately not awarded time. J0841+0101, on the other hand, was targeted in a separate, joint XMM-Newton-NuSTAR follow-up study\footnote{For this joint observation, we used the combined count rates found by Chandra to determine count rates for XMM via PIMMs, assuming heavy absorption, and to generate fake NuSTAR spectra comprising several hundred counts using the NuSTAR responses.} aimed at confirming the reported Fe K$\alpha$ \citep{pfeifle2019a} and placing more stringent constraints on the column density. The observations were conducted with the targets at the aimpoint, with total exposures times ranging from 29.9 ks to 49.7 ks. Details on the observations are shown in Table~\ref{table:nustarobs}. Observations of three other mergers in our sample, J0849+1114, J1301+2911 (NGC\,4922), and J1356+1822 (Mrk\,463), were previously published and studied by \citet{pfeifle2019b}, \citet{ricci2017MNRAS}, and \citet{yamada2018}, respectively; rather than reanalyzing these systems, we refer to the results of these previous works where relevant in this work. The \nustar{} observations were processed using the \nustar{} Data Analysis Software \citep[\textsc{nustardas},][]{nustardas}\footnote{https://heasarc.gsfc.nasa.gov/docs/nustar/analysis/} v0.4.7 package available in \textsc{heasoft} v6.27 \citep{heasoft}, along with latest \textsc{caldb} version and the clock correction file at the time of reprocessing. Level two data products were created using \textsc{nupipeline} (v0.4.7), and specific processing choices to account for the South Atlantic Anomaly (SAA) were made based upon the provided background light curves for focal plane modules A and B (FPMA and FPMB); the \textsc{tentacle yes} option was used in cases where the background was not stable over the duration of the observation (as per the \textsc{nustardas} manual). The \textsc{dmcopy} tool contained within the Chandra Interactive Analysis of Observations (\textsc{ciao}) software package \citep{fruscione2006} was used to create energy filtered images for the $3-10$\,keV (soft), $10-24$\,keV (hard), and $3-24$\,keV (full) \nustar{} energy bands. As discussed below in Section 3, counts for specific energy bands were extracted from these energy filtered science images. Where spectral extraction was relevant, we used the \textsc{nuproducts} script to extract the Stage 3 data products using the source and background regions described in Section~\ref{sec:sourcedet}, yielding source and background spectra as well as the relevant response files extracted across the 3.0-78.0 keV energy range. We grouped the FPMA and FPMB spectra by 1\,count per bin using the \textsc{heasoft} \textsc{grppha} command in order to fit the spectra in \textsc{xspec} using Cash statistics \citep{cash1979}, which is more appropriate than $\chi^2$ statistics given the low number of counts within the spectra.

\subsection{\xmm{} Observations}
\label{sec:xmmobs}
J0841+0101 was observed on-axis by \xmm{} (HEW: $\sim17$\arcsec{}) on 2018 May 10 for a total of 32 ks (see Table~\ref{table:nustarobs} for details). We reprocessed the observation using SAS v20.0.0, cleaning the event files of bad pixels, bad patterns, and background flares. We generated energy filtered science images for each of the EPIC cameras using the SAS \textsc{evselect} command. Source spectra were generated using the \textsc{evselect} command using a 30'' radius source aperture and a 1' radius background aperture (placed in a source free region on the same CCD); response files were generated using the \textsc{rmfgen} and \textsc{arfgen} commands. We grouped the spectra using \textsc{grppha} at 1 count per bin in order to use Cash statistics \citep{cash1979} during spectral fitting in \textsc{xspec}.

\subsection{\chandra{} Observations}
\label{sec:chandraobs}
In addition to the \nustar{} and \xmm{} observations of J0841+0101, we also retrieved the archival \chandra{} ACIS-S observations for use during the spectral analysis. J0841+0101 was observed for 20\,ks on 2012 February 25 and for 23\,ks on 2016 January 10, and these observations were examined and published by \citet{comerford2015},  \citet{pfeifle2019a}, and \citet{foord2020}. We reprocessed the observations using \textsc{ciao} v.4.14 and \textsc{caldb} v.4.9.7 using the \textsc{chandra\_repro} script. Energy filtered images were again created using \textsc{dmcopy} within \textsc{ciao}. Spectra were extracted using the \textsc{specextract} script, and we used the 1.5\arcsec{} source and associated background aperture choices from \citet{pfeifle2019a} to extract the source and background spectra. As in the case of the \nustar{} and \xmm{} data, spectra were grouped by 1 count per bin.

\section{Data Analysis}

\subsection{X-ray Photometric Analysis}
\subsubsection{Source Detection and Photometry}
\label{sec:sourcedet}
\nustar{} and \xmm{} source aperture positions were chosen using the \textsc{wavdetect} tool within the \textsc{ciao} package \citep{fruscione2006} and the use of the ``centroid'' feature in DS9 when necessary; for non-detected systems, we placed apertures at the SDSS positions of the galaxy merger. Source apertures 45'' in radius ($\sim64$\% enclosed energy fraction) were used to extract NuSTAR spectra and counts for each source while background annuli of inner radius 90'' and outer radius 150'' were used to extract background spectra and counts. An inner radius of 90'' ensures minimal contribution from the source. The effective area of \nustar{} drops off dramatically after $24-30$ keV, hindering the utility of the higher energy bands; we limit our photometric investigation to the $3-24$ keV energy range for the \nustar{} data, although for the spectroscopic investigation of J0841+0101, we use the $3.0-30$\,keV range for the \nustar{} data. %Most of our sources do not show emission beyond 24 keV relative to the background, and moreover %The filtered images and the source apertures described above in Section 2 were used to extract the source photometry.

For \nustar{}, we extracted counts from the energy filtered images using the \textsc{dmextract} package within \textsc{ciao}. We assumed Gaussian statistics in the case of detected sources with greater than 20 counts, computing the source error as $\sqrt{\rm{N}}$ (where N is the number of counts), and generated background subtracted counts after normalizing the background counts to the size of the source region. We derived formal statistical significance levels using the background subtracted counts and required a significance threshold of $3\sigma$ for a source to be considered formally detected within a particular energy band. As an additional check, the logarithm of the binomial no-source probability \citep{lansbury2014} was computed for each system, which can be particularly important for identifying weak sources that do not necessarily meet formal detection thresholds. The minimum threshold for the no-source statistic is log($P_{\rm{B}})\leq-2.7$, meaning that if log($P_{\rm{B}})\leq-2.7$, the gross counts within an aperture are unlikely to be purely the result of a background fluctuation. For instances of non-detections in the \nustar{} imaging, we instead computed the $3\sigma$ (99.7\% confidence level) upper limits for the net source counts following the Bayesian method of \citet{kraft1991}, which takes into account the gross measured counts in the source and background regions. 

Table~\ref{table:nustarphotometry} shows the FPMA and FPMB background subtracted photometry for all of the dual AGN candidates observed by \nustar{}; no correction for the encircled energy fraction was made to these net counts. J0122+0100, J1221+1137, and J1306+0735 were not formally detected or identified using the binomial source statistic in the \nustar{} imaging. J0841+0101 was clearly detected by both \nustar{} and \xmm{}.

\subsubsection{\nustar{} Flux Calculations}
\label{sec:nustarfluxes}
For each of the non-detected dual AGN candidates observed by \nustar{}, we converted the observed count rates to observed fluxes (or flux upper limits) in each of the $3-24$\,keV, $3-10$\,keV, $10-24$\,keV, and $2-10$\,keV (in order to directly compare to our previous work with \chandra{}, i.e. \citealp{pfeifle2019b}) energy bands using the \chandra{} \textsc{pimms} toolkit. To accomplish this, we combined the count rates from FPMA and FPMB for each energy band, and then scaled the combined count rates to what would have been derived from an aperture enclosing 50\% of the enclosed energy (EEF), as required before inputting the count rates into \textsc{pimms}. When calculating fluxes, we assumed a simple power law with $\Gamma=1.8$ \citep[e.g.,][]{mushotzky1993,ricci2017bass}, and we provided \textsc{pimms} the EEF scaled fluxes, the SDSS spectroscopic redshifts of the merging systems, and the Galactic $N_{\rm{H}}$ along the line-of-sight, which we retrieved from the \textit{Swift} Galactic \nh{} calculator\footnote{https://www.swift.ac.uk/analysis/nhtot/}. The resulting fluxes are tabulated in Table~\ref{table:nustarfluxes}. Note for the case of J0841+0101, we performed detailed spectral fitting and therefore did not need to perform this simple count rate to flux conversion.

\begin{figure*}
    \centering
    \includegraphics[width=0.8\linewidth]{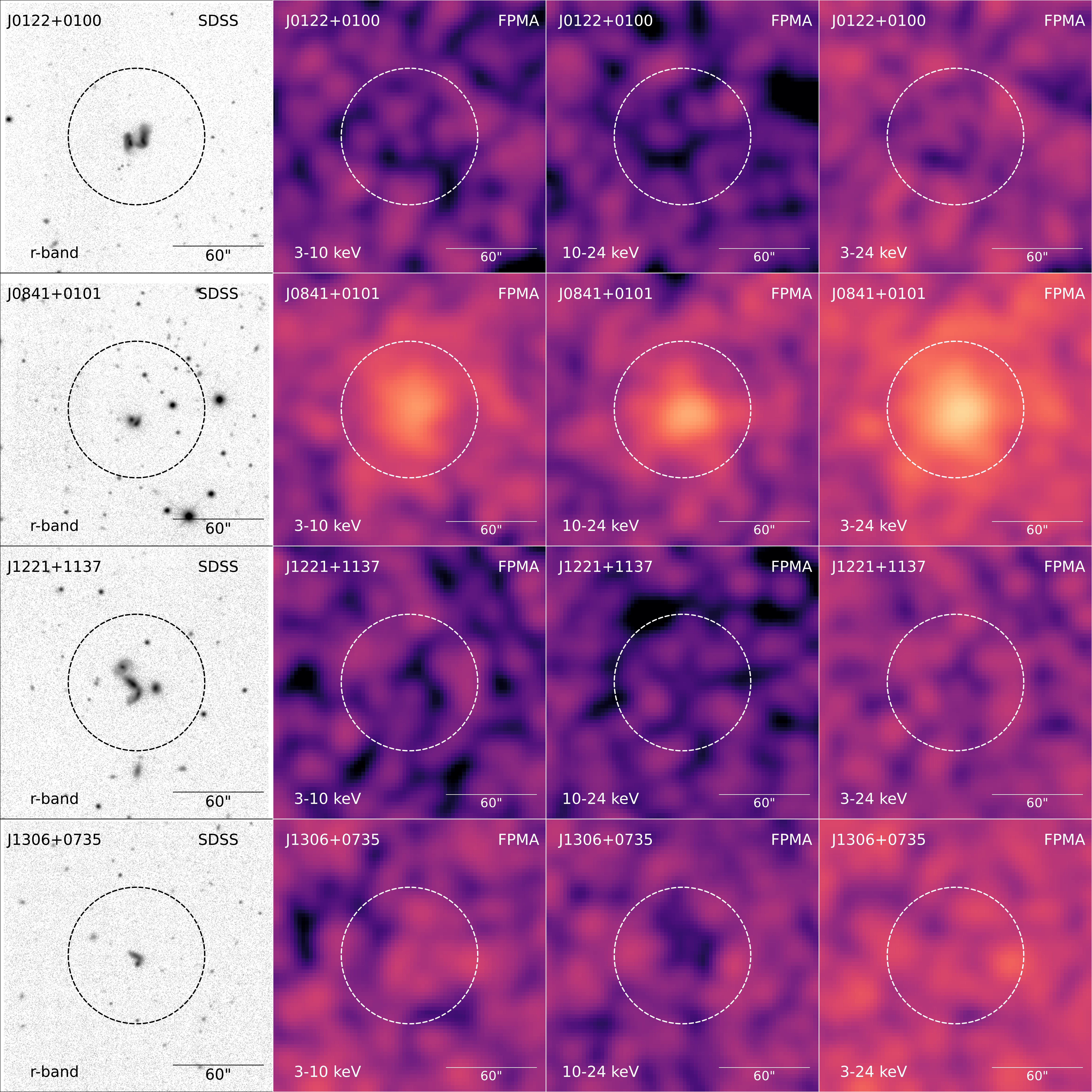}
    \caption{SDSS r-band and \nustar{} FPMA imaging for (top to bottom) J0122+0100, J0841+0101, J1221+1137, and J1306+0735. Left to right: SDSS r-band, \nustar{} FPMA $3-10$\,keV, $10-24$\,keV, and $3-24$\,keV bands. We smoothed the X-ray images with a three-pixel Gaussian kernel and use the perceptually uniform sequential color map `magma' in \textsc{matplotlib}. Dashed $45$\arcsec{} radius circles represent the source extraction regions. The scale bars in the lower right corner of each panel indicates an angular size of 60''.}
    \label{fig:nustarimaging}
\end{figure*}

\subsubsection{Indirect Column Density Estimation}
\label{sec:nhest}
As mentioned above in Section~\ref{sec:sourcedet} and discussed below in Section~\ref{sec:photresults}, J0122+0100, J1221+1137, and J1306+0735 were not detected by \nustar{}, hindering our ability to constrain the column densities along the line-of-sight. To circumvent this issue, we indirectly derived lower limits on the column density for each of these dual AGN candidates by comparing the ratio of the hard X-ray flux upper limits and the expected, unobscured hard X-ray fluxes ($F_{\rm{10-24\,keV,\,upper\,limit}}$/$F_{\rm{10-24\,keV,\,expected}}$) to sets of attenuation curves \citep[e.g.,][]{ricci2015} generated in \textsc{xspec}. This technique was used in \citet{pfeifle2022b}, and we briefly describe the process here:
\begin{itemize}
    \item We first began by calculating the hard X-ray $10-24$\,keV flux upper limit for each dual AGN candidate as described in Section~\ref{sec:nustarfluxes}.
    \item Then we calculated the expected, intrinsic hard X-ray fluxes. First, we estimated the expected $2-10$\,keV flux using the \citet{asmus2015} relation between the $2-10$\,keV and 12\,$\rm{\mu m}$ fluxes, where we used the AGN 12\,$\rm{\mu m}$ flux derived from our latest SED fitting, which we describe in Appendix~\ref{sec:seds}. With the expected, intrinsic $2-10$\,keV flux in hand, we then converted this to the $10-24$\,keV flux using a scale factor derived from a power law model in \textsc{xspec} assuming $\Gamma=1.8$.
    \item We then took the ratio between the 10-24\,keV flux upper limit from \nustar{} and the expected 10-24\,keV flux.
    \item Next we established an attenuation curve for each merger system. This was generated with a model in \textsc{xspec} that includes a primary power law, photoelectric absorption, Compton scattering, reprocessed emission from a torus, and Thomson scattering, expressed in \textsc{xspec} as:
    $(\textsc{f}\times\textsc{cutoffpl}) +(\textsc{tbabs}\times\textsc{cabs}\times\textsc{cutoffpl}) +\textsc{Borus}$.
    This model assumed $\Gamma=1.8$ \citep{mushotzky1993,ricci2017bass}, a scattering fraction of 0.05\% ($f=0.005$), and the model was stepped through column densities of log($\nhm{}$)=22-25.4 in increments of $\Delta\rm{log}(\nhm{})=0.2$ in order to build the attenuation curve as a function of column density. During this step, we built two curves per system, one for each choice of covering factor: C=0.5 and C=0.99, where 0.99 may be more appropriate in merging systems where the obscuring material may take up a large fraction of the sky as seen by the AGN \citep{ricci2017MNRAS,ricci2021,yamada2021}. These curves were normalized by the `intrinsic' $10-24$\,keV flux of the model measured when $\rm{log}(\nhm{})=22.0$. %An example of these curves is shown in Figure~\ref{fig:attencurves}.
\end{itemize}
Once the attenuation curves were generated for each merger, and the ratio of $F_{\rm{10-24\,keV,\,upper\,limit}}$/$F_{\rm{10-24\,keV,\,expected}}$ was known, we interpolated the flux ratio between values in the attenuation curves to derive the column densities required for the observed X-ray suppression. We list the derived column densities in Table~\ref{table:nustarfluxes} alongside the \nustar{} fluxes derived in Section~\ref{sec:nustarfluxes}. 

\begin{table*}
\centering{
\caption{\nustar{} Photometry}
\label{table:nustarphotometry}
\begin{tabular}{cccccccc}
\hline
\hline
\noalign{\smallskip}
\noalign{\smallskip}
System & FPM & $\alpha$ & $\delta$ & $z$ & \multicolumn{3}{c}{Net Counts} \\ 
 & & & & & $3-24$\,keV & $3-10$\,keV & $10-24$\,keV \\
(1) & (2) & (3) & (4) & (5) & (6) & (7) & (8) \\
%& Aperture RA & Aperture Dec 
\noalign{\smallskip}
\noalign{\smallskip}
\hline
\noalign{\smallskip}
J0122+0100  & A &  1\hr{}22\min{}17.\sec{}839 & +1\deg{}00\min{}30.\sec{}336 & 0.05546 & $<24.3$  & $<21.0$ & $<17.0$  \\
            & B &  1\hr{}22\min{}17.\sec{}839 & +1\deg{}00\min{}30.\sec{}336 & 0.05546 & $<50.6$  & $<30.7$ & $<35.8$  \\
J0841+0101  & A &  8\hr{}41\min{}34.\sec{}879 & +1\deg{}02\min{}03.\sec{}041 & 0.11060 & $296.7\pm23.9$ & $151.2\pm17.2$ & $148.9\pm16.6$ \\
%296.7\pm23.9 & 151.2\pm17.2 & 148.9\pm16.6
            & B &  8\hr{}41\min{}35.\sec{}306 & +1\deg{}01\min{}54.\sec{}559 & 0.11060 & $331.5\pm25.4$ & $176.3\pm19.2$ & $154.5\pm16.7$ \\
%331.5\pm25.4 & 176.3\pm19.2 & 154.5\pm16.7
J1221+1137  & A & 12\hr{}21\min{}04.\sec{}845 & 11\deg{}37\min{}53.\sec{}570 & 0.06820 & $<22.5$ & $<20.2$ & $<16.1$   \\
            & B & 12\hr{}21\min{}04.\sec{}845 & 11\deg{}37\min{}53.\sec{}570 & 0.06820 & $<37.6$ & $<30.3$ & $<25.4$   \\
J1306+0735  & A & 13\hr{}06\min{}53.\sec{}601 & +7\deg{}35\min{}18.\sec{}850 & 0.11111 & $<33.8$ & $<32.6$ & $<20.0$   \\
            & B & 13\hr{}06\min{}53.\sec{}601 & +7\deg{}35\min{}18.\sec{}850 & 0.11111 & $<31.7$ & $<33.6$ & $<17.7$   \\

\noalign{\smallskip}
\hline
\end{tabular}
}
%\end{center}
\tablecomments{\nustar{} photometry for the $3-24$\,keV (full), $3-10$\,keV (soft), or $10-24$\,keV (hard) energy bands. Col. 1-2: System name and FPM. Col. 3-5: right ascension, declination, and redshift. Col. 6-8: full band, soft band, and hard band counts. No correction was made for the enclosed energy fraction in this table.} 
\end{table*}

\subsection{Spectroscopic Analysis}
\label{sec:specanalysis}
J0841+0101 was the only source formally detected by \nustar{} with statistical significance, and here we have used the new \nustar{} and \xmm{} observations, in conjunction with the archival \chandra{} observations, to study the X-ray spectral properties in the 0.1-30\,keV band. We limited our analysis to the 0.1-30 keV energy band due to the drop in effective area of \nustar{} beyond $\sim30$\,keV. As we moved from simpler models to more complex models, for each spectral model component with one free parameter (introducing one new degree of freedom), the C-stat value for the fit was required to change such that $\rm{Cstat}_{\rm{old}}-\rm{Cstat}_{\rm{new}}>2.71$ to be considered a statistically significant improvement to the fit \citep[e.g.,][]{tozzi2006,brightman2014}, and each new spectral model was visually inspected to ensure a proper fit has been found. 

We began our fitting process with a simple phenomenological power law model that accounts for Galactic $N_{\rm{H}}$ and includes photoelectric absorption, Compton scattering, and Thompson scattering (in the form of a scattered power law), given in \textsc{xspec} as: \textsc{tbabs$\times$(f$\times$cutoffpl+tbabs$\times$cabs$\times$cutoffpl)}. The normalization and photon index of the scattered component were tied to that of the intrinsic power law. Residuals leftover after this initial fit suggested the presence of two thermal components in the 0.1-1\,keV band, which we modeled with two separate \textsc{apec} components; iteratively adding and fitting the spectra with these additional components yielded $\Delta$C-stat=81.8 and $\Delta$C-stat=33.1. To account for excess emission above 10\,keV, we included a reflection component off of a cold slab, described by the \textsc{pexrav} model in Xspec ($\Delta$C-stat=18.5); we set $\rm{R}<0$ so that this component represented only a reflection component and did not include an intrinsic power law component. We assume solar abundances and an inclination angle of 60\deg{} for each fit; these parameters were not free to vary. Finally, strong excess emission was observed near 6.4\,keV (in the source rest frame), indicative of an Fe K$\alpha$ emission line \citep[originally identified by][using \chandra{} X-ray observations]{pfeifle2019a}; we modeled this emission using a Gaussian line (\textsc{zgauss}) with the line centroid fixed to 6.4\,keV and the width fixed to 0.01\,keV to model only the narrow component of the line ($\Delta$C-stat=20.9). The best fitting phenomenological model is given in \textsc{xspec} as:
\textsc{tbabs$\times$(apec+apec+f$\times$cutoffpl} \textsc{+ztbabs$\times$cabs$\times$cutoffpl+pexrav+zgauss)}.

We followed this procedure again but this time employed the physically-motivated and self-consistent torus model, \textsc{borus} \citep{balokovic2018}, to account for reprocessing of the intrinsic AGN emission by an obscuring torus instead of using a \textsc{pexrav} component. There is an important distinction between the phenomenological model and this physically-motivated model: \textsc{borus} provides a parameter for the torus column density, which can be different than the line-of-sight (LOS) column density. For simplicity, we began the fitting process using a model that accounts for the torus column density, an absorbed power law (with LOS column density tied to that of the torus column density for a simpler geometric case) that accounts for photoelectric absorption and Compton scattering as well as Thompson scattering, given in 
\textsc{xspec} as: \textsc{tbabs$\times$(f$\times$cutoffpl+ztbabs$\times$cabs$\times$cutoffpl +borus)}. We assume solar abundances, a torus half opening angle of 60\deg{} (corresponding to a covering factor of 50\%), and an inclination angle of 70\deg{} for each fit; these parameters were not free to vary. We also tied the power law photon indices and normalizations of the power law models and the \textsc{borus} model together. Like in the case of the phenomelogical model, we found that the addition of two \textsc{apec} components ($\Delta$C-stat=61.8 and $\Delta$C-stat=34.8) statistically and visually improved the fit. Furthermore, we found that untying the LOS and torus column densities resulted in a statistically significant improvement to the fit ($\Delta$C-stat=11.7). The final best fitting model for this physically-motivated case was given by: \textsc{tbabs$\times$(apec+apec+f$\times$cutoffpl} \textsc{+ztbabs$\times$cabs$\times$cutoffpl+borus)}. \textsc{borus} self consistently models the Fe K$\alpha$ emission line, so there was no need to include a Gaussian emission line component to the model. 

Throughout the fitting process, we froze the redshift to the spectroscopic redshift given by SDSS and froze the Galactic $N_{\rm{H}}$ to that given by the \textit{Swift} Galactic $N_{\rm{H}}$ calculator \citep{willingale2013}. A constant was prepended to each model and left free to vary within $\pm30$\% of the first data instance loaded into \textsc{xspec} to account for inter-detector sensitivity; during the fitting process, these constants were monitored to ensure that they did not vary by more than roughly $\pm15\%$ (all but one constant remained $<10$\%). 

\begin{table*}
\begin{center}
\caption{X-ray Properties for the \nustar{} Non-Detected AGNs}
\label{table:nustarfluxes}
\begin{tabular}{cccccccc}
\hline
\hline
\noalign{\smallskip}
\noalign{\smallskip}
System & \multicolumn{4}{c}{\nustar{}} Observed Flux & \chandra{} Flux & Covering Factor & log($N_{\rm{H}}/\rm{cm}^{-2}$) \\ 
 & \multicolumn{4}{c}{($10^{-13}$ erg cm$^{-2}$ s$^{-1}$)} & ($10^{-13}$ erg cm$^{-2}$ s$^{-1}$) & \\
 & $3-24$\,keV & $3-10$\,keV & $10-24$\,keV & $2-10$\,keV & $2-10$\,keV &  & \\
(1) & (2) & (3) & (4) & (5) & (6) & (7) & (8)\\
%& Aperture RA & Aperture Dec 
\noalign{\smallskip}
\noalign{\smallskip}
\hline
\noalign{\smallskip}
J0122+0100 & $<0.895$ & $<0.434$ & $<1.19$ & $<0.558$ & $0.14\pm0.02$ & 0.5 (0.99) & $>$24.7 ($>$24.9)\\
%J0841+0101 & $12.61\pm0.68$ & $4.67\pm0.35$ & $11.42\pm0.95$ & $5.99\pm0.45$ & \dots & \dots \\
J1221+1137 & $<1.81$ & $<1.07$ & $<2.36$ & $<1.37$ & $0.08\pm0.03$ & 0.5 (0.99) & $>$24.4 ($>$24.6)\\
J1306+0735 & $<1.13$ & $<0.80$ & $<1.23$ & $<1.03$ & $0.08\pm0.01$ & 0.5 (0.99) & $>$24.1 ($>$24.3) \\
\noalign{\smallskip}
\hline
\end{tabular}
\end{center}
\tablecomments{\nustar{} fluxes (or flux upper limits) derived using \textsc{pimms} assuming a power law index of $\Gamma=1.8$ for the $3-24$\,keV, $3-10$\,keV, $10-24$\,keV, and $2-10$\,keV energy bands. Col. 1: System name. Col. 2-5: fluxes for the $3-24$\,keV, $3-10$\,keV, $10-24$\,keV, and $2-10$\,keV energy bands. Col. 6: observed $2-10$\,keV flux from \citet{pfeifle2019a} where the values reported here are the sums of the X-ray fluxes of the nuclei in each merger. Col. 7: choice of covering factor (see Section~\ref{sec:nhest}). Col. 8: Column density lower limits derived using the attenuation curves and flux ratios described in Section~\ref{sec:nhest}; The values in parenthesis are calculated assuming a covering factor of 0.99 rather than 0.5. Note: we do not provide a column density lower limit for J0841+0101 since we were able to fit the spectra for that system.
}
\end{table*}

\begin{table*}
\begin{center}
\caption{X-ray Luminosities for the \nustar{} Non-Detected AGNs}
\label{table:nustarluminosities}
\begin{tabular}{cccccc}
\hline
\hline
\noalign{\smallskip}
\noalign{\smallskip}
System & \multicolumn{4}{c}{\nustar{}} Observed Luminosity & \chandra{} Observed Luminosity \\ 
 & \multicolumn{4}{c}{($10^{42}$ erg s$^{-1}$)} & ($10^{42}$ erg s$^{-1}$) \\
 & $3-24$\,keV & $3-10$\,keV & $10-24$\,keV & $2-10$\,keV & $2-10$\,keV \\
(1) & (2) & (3) & (4) & (5) & (6) \\
%& Aperture RA & Aperture Dec 
\noalign{\smallskip}
\noalign{\smallskip}
\hline
\noalign{\smallskip}
J0122+0100 & $<0.656$ & $<0.318$ & $<0.873$ & $<0.409$ & $0.11\pm0.02$\\
J1221+1137 & $<2.04$ & $<1.21$ & $<2.66$ & $<1.55$ & $0.09\pm0.03$\\
J1306+0735 & $<3.59$ & $<2.55$ & $<3.90$ & $<3.28$ & $0.24\pm0.04$\\
\noalign{\smallskip}
\hline
\end{tabular}
\end{center}
\tablecomments{Luminosities (or upper limits) for the $3-24$\,keV, $3-10$\,keV, $10-24$\,keV, and $2-10$\,keV energy bands; see Section~\ref{sec:nustarfluxes} and Table~\ref{table:nustarfluxes} for the calculation of the fluxes. Col. 1: System name. Col. 2-5: NuSTAR luminosities for the $3-24$\,keV, $3-10$\,keV, $10-24$\,keV, and $2-10$\,keV energy bands. Col. 6: the observed $2-10$\,keV Chandra luminosity reported in \citet{pfeifle2019a}, where the luminosities quoted here are the sums of the luminosities of the nuclei in each merger.}

\end{table*}

\section{Results}
\subsection{Photometric Results}
\label{sec:photresults}
Of the four newly imaged dual AGN candidates, three (J0122+0100, J1221+1137, and J1306+0735) were not detected by \nustar{} ($<3\sigma$) and the binomial no-source statistics were inconclusive. The source positions and upper limits on the counts for these three systems are reported in Table~\ref{table:nustarphotometry}. J0841+0101, on the contrary, is well detected by \chandra{}, \xmm{}, and \nustar{} in every energy band explored, and the counts from the \nustar{} observation are reported in Table~\ref{table:nustarphotometry}. Using the count rates and upper limits, we calculated the $3-24$\,keV, $3-10$\,keV, $10-24$\,keV, and $2-10$\,keV fluxes and flux upper limits for all four of the newly observed systems, assuming a power law index of $\Gamma=1.8$ (see Section~\ref{sec:nustarfluxes}). The non-detections in J0122+0100, J1221+1137, and J1306+0735 suggest that there are no $10-24$\,keV X-ray emitting AGNs in excess of $1.19\times10^{-13}$, $2.36\times10^{-13}$, and $1.23\times10^{-13}$ erg s$^{-1}$ cm$^{-2}$, respectively, in these systems. 

To offer a direct comparison to our previous work with \chandra{}, we tabulated the $2-10$\,keV fluxes derived from \nustar{} in Table~\ref{table:nustarphotometry}. The previous \chandra{} observations provided far more stringent measurements of the $2-10$\,keV fluxes in these mergers \citep[$\approx1.4\times10^{-14}$\,erg\,cm$^{-2}$\,s$^{-1}$ for J0122+0100, $\approx8.4\times10^{-15}$\,erg\,cm$^{-2}$\,s$^{-1}$ for J1221+1137, and $\approx7.5\times10^{-15}$\,erg\,cm$^{-2}$\,s$^{-1}$ for J1306+0735,][]{pfeifle2019a}; in each of the three non-detection cases presented here, the $2-10$\,keV flux upper limits derived from \nustar{} are well in excess of the measured $2-10$\,keV fluxes reported in \citet{pfeifle2019a}, so it is not surprising that the mergers remained undetected in the softer $3-10$\,keV \nustar{} energy band. To better illustrate this point, we display the intrinsic AGN 12\,$\mu$m luminosities (derived via SED fitting; see Appendix~\ref{sec:seds}) and the $2-10$\,keV luminosities derived via Chandra and NuSTAR for the sub-sample examined here in Figure~\ref{fig:lx_v_l12} along with the Chandra $2-10$\,keV luminosities for the remainder of the sample \citep[from][]{pfeifle2019a}.

Without spectra for J0122+0100, J1221+1137, and J1306+0735, we indirectly estimated lower limits on the column densities along the line-of-sight as outlined in Section~\ref{sec:nhest}. The $10-24$\,keV flux upper limits derived from \nustar{} imply column densities of log($\nhm{}$) $\geq$ 24.9, 24.8, and 24.6 (assuming a covering factor of C = 0.99), respectively, assuming the non-detection in the $10-24$\,keV band is due purely to obscuration. These column density lower limits are roughly consistent with what we found in \citet{pfeifle2019a}, where the differences are likely due to the different methods used to estimate the column densities. 

\subsection{Spectroscopic Results}
Our broad band (0.1-30 keV) joint \nustar{}, \xmm{}, and \chandra{} spectroscopic analysis for J0841+0101 yielded direct constraints on the AGN X-ray properties. In Table~\ref{table:specanalysis} we present the best fitting parameters for our two best fitting models, described in Section~\ref{sec:specanalysis}, one phenomenological model which we refer to as the ``\textsc{pexrav}'' model, and one physical torus model which we refer to as the ``\textsc{borus}'' model. Each of these best fitting models feature two thermal components in the 0.1-1 keV band ($T_1 = 0.17_{-0.06}^{+0.03}$ keV and $T_2 = 0.90_{-0.11}^{+0.13}$ keV for the \textsc{pexrav} model, or $T_1=0.16_{-0.06}^{+0.04}$ keV and $T_2=0.90_{-0.12}^{+0.14}$ keV for the \textsc{borus} model), scattered emission ($f=2.7_{-2.2}^{+16.1}$\% or $0.7_{-0.3}^{0.4}$\%), photon indices typical of Seyfert 2 AGNs ($1.9_{-0.4}^{+0.3}$ for the \textsc{pexrav} model or $2.0_{-0.2}^{+0.2}$ for the \textsc{borus} model), components for reflection and reprocessing ($R=-2.3_{-6.9}^{+1.6}$ for the \textsc{perxrav} model), iron K$\alpha$ fluorescent emission lines (equivalent width of 0.3 keV in the case of the \textsc{pexrav} model), and line-of-sight absorption. The key difference between these models is how the line-of-sight obscuration is handled: for the \textsc{borus} model, we allow the torus column density to vary independently of the line-of-sight obscuration, and the best-fitting model finds different column densities for each: a torus average column density (given in log units, as is provided by the model) of log($\nhm{}) = 23.0_{-0.1}^{0.1}$ and a line-of-sight column density \nh{} = $121_{-18}^{+20}\times10^{22}\,\rm{cm}^{-2}$. The \textsc{pexrav} model, on the other hand, allowed for only a single column density along the line-of-sight, which yielded \nh{} = $80.1_{-52.8}^{+44.9}\times10^{22}$\,cm$^{-2}$; nonetheless, the column density derived using the \textsc{pexrav} model is consistent with the line-of-sight column density derived using the \textsc{borus} model within the quoted error bounds. 

These column densities are higher than those derived using the \chandra{} data alone in \citet{pfeifle2019a}, though there is significant overlap when considering the error bounds in Table~\ref{table:specanalysis}: specifically, the LOS column density of the \textsc{pexrav} model is consistent within the error bounds with that found in \citet{pfeifle2019a}, while the torus column density of the \textsc{borus} model is slightly smaller than what was found in \citet{pfeifle2019a}. Nonetheless, the LOS column density found with the \textsc{borus} model does imply a significantly higher column density than was previously reported. The power law index values found in this work are better constrained than with the \chandra{} data alone in \citet{pfeifle2019a}, but the scattered fractions determined here are consistent with the values found in \citet{pfeifle2019a}. We also confirm the presence of the iron Fe K$\alpha$ line previously reported in \citet{pfeifle2019a}, though with a smaller equivalent width \citep[$0.26^{+0.15}_{-0.07}$ keV vs. 0.75 keV in][]{pfeifle2019a}. The refinement of the X-ray properties in this work relative to previous works is not surprising; the addition of the \xmm{} spectra with higher signal to noise in the 0.1-10\,keV band in concert with access to the hard energy 3-30 keV band via \nustar{} yielded better constraints on the AGN X-ray properties. 

The analysis presented here comes with a clear caveat: we assume there is only a single hard X-ray emitting AGN in this merging system; in reality, if the second galaxy hosts a heavily Compton-thick AGN that emits strongly in the hard X-rays, this AGN would contribute appreciably to the observed hard X-ray spectrum, and thus the derived X-ray spectral properties would describe the sum of the spectra of the two sources rather than the specific properties attributable to one AGN or the other; at the present time, the spectra are not of sufficient quality to attempt such a deconvolution. While the spectroscopic analysis here does not provide evidence for or against the dual AGN scenario for J0841+0101 \citep[it is unclear if this is a dual AGN; see][who found that the system is consistent with a single X-ray emitting AGN using the BAYMAX tool]{foord2020}, it nonetheless represents a clear case of heavy absorption induced by a merger. Our X-ray spectroscopic results showing the presence of a strong soft X-ray emission component are consistent with the analysis presented in \citet{foord2020}, who concluded that J0841+0101 comprises a single resolved X-ray source surrounded by diffuse and extended emission.

\begin{table*}
\begin{center}
\caption{Spectral Fitting Results for J0841+0101}
\label{table:specanalysis}
\begin{tabular}{ccccccccccccc}
\hline
\hline
\noalign{\smallskip}
\noalign{\smallskip}
Model & C-stat & d.o.f & $\Gamma$ & $N_{\rm{H}, Torus}$ & $N_{\rm{H}, LOS}$ & R & $f_{\rm{S}}$ & T$_1$ & T$_2$ & Fe K$\alpha$ EW \\
 &  &  &  & (log[cm$^{-2}$]) & ($10^{22}$ cm$^{-2}$) &  & (\%) & (keV) & (keV) & (keV) \\%& (erg\,s$^{-1}$) & (erg\,s$^{-1}$) \\
(1) & (2) & (3) & (4) & (5) & (6) & (7) & (8) & (9) & (10) & (11) \\%& (12) & (13)\\
\noalign{\smallskip}
\noalign{\smallskip}
\hline
\noalign{\smallskip}
Pexrav & 1738.35 & 1914 & $1.9_{-0.4}^{+0.3}$ & \dots & $80.1_{-52.8}^{+44.9}$ & $-2.3_{-6.9}^{+1.6}$ & $2.7_{-2.2}^{+16.1}$ & $0.17_{-0.06}^{+0.03}$ & $0.90_{-0.11}^{+0.13}$ & $0.26^{+0.15}_{-0.07}$ \\
Borus & $1736.80$ & $1914$ & $2.0_{-0.2}^{+0.2}$ & $23.0_{-0.1}^{+0.1}$ & $121_{-18}^{+20}$ & \dots & $0.7_{-0.3}^{+0.4}$ & $0.16_{-0.06}^{+0.04}$ & $0.90_{-0.12}^{+0.14}$ & \dots \\
\noalign{\smallskip}
\hline
\end{tabular}
\end{center}
\tablecomments{J0841+0101 spectral fitting results. Col. 1: Model choice. Col. 2-3: C-stat and degrees of freedom of the model. Col. 4-6: Photon index, torus column density, and line-of-sight column density. Col. 7-8: reflection coefficient, scattered fraction. Col. 9-10: temperatures for the two apec components. Col. 11-13: iron K$\alpha$ line equivalent width. %, the intrinsic $2-10$\,keV luminosity, and the intrinsic $3-24$\,keV luminosity.
}
\end{table*}

\begin{table*}
\begin{center}
\caption{Absorbed and Unabsorbed Fluxes for J0841+0101}
\label{table:speclums}
\begin{tabular}{ccccccccc}
\hline
\hline
\noalign{\smallskip}
\noalign{\smallskip}
Model & \multicolumn{2}{c}{$F_{\rm{2-10\,keV}}$} & \multicolumn{2}{c}{$F_{\rm{3-24\,keV}}$} & \multicolumn{2}{c}{$L_{\rm{2-10\,keV}}$} & \multicolumn{2}{c}{$L_{\rm{3-24\,keV}}$}\\
& \multicolumn{2}{c}{($10^{-13}$\,erg\,cm$^{-2}$\,s$^{-1}$)} & \multicolumn{2}{c}{($10^{-13}$\,erg\,cm$^{-2}$\,s$^{-1}$)} & \multicolumn{2}{c}{($10^{42}$\,erg\,s$^{-1}$)} & \multicolumn{2}{c}{($10^{42}$\,erg\,s$^{-1}$)}\\
& Obs. & Intr. & Obs. & Intr. & Obs. & Intr. & Obs. & Intr. \\
%& (erg\,s$^{-1}$) & (erg\,s$^{-1}$) \\
(1) & (2) & (3) & (4) & (5) & (6) & (7) & (8) & (9)\\%& (12) & (13)\\
\noalign{\smallskip}
\noalign{\smallskip}
\hline
\noalign{\smallskip}
Pexrav &  $2.03^{+0.29}_{-0.62}$ & 9.36 & $9.28^{+0.69}_{-2.12}$ & 18.0 & $6.38^{+0.91}_{-1.95}$ & 29.43 & $29.19^{+2.17}_{-6.67}$ & 56.61\\
Borus &  $2.06^{+0.22}_{-0.94}$ & 37.8 & $4.77^{+0.61}_{-1.95}$ & 51.6 & $6.48^{+0.69}_{-2.96}$ & 118.9 & $15.0^{+1.92}_{-6.13}$ & 162.3 \\
\noalign{\smallskip}
\hline
\end{tabular}
\end{center}
\tablecomments{J0841+0101 spectroscopically-derived fluxes. Col. 1: Model choice. Col. 2-3: the absorbed and intrinsic $2-10$\,keV flux. Col 4-5: the absorbed and intrinsic $3-24$\,keV flux. Col. 6-7: the absorbed and intrinsic $2-10$\,keV luminosity. Col 8-9: the absorbed and intrinsic $3-24$\,keV luminosity.}
\end{table*}

\begin{figure}
    \centering
    \subfloat{\includegraphics[width=1.0\linewidth]{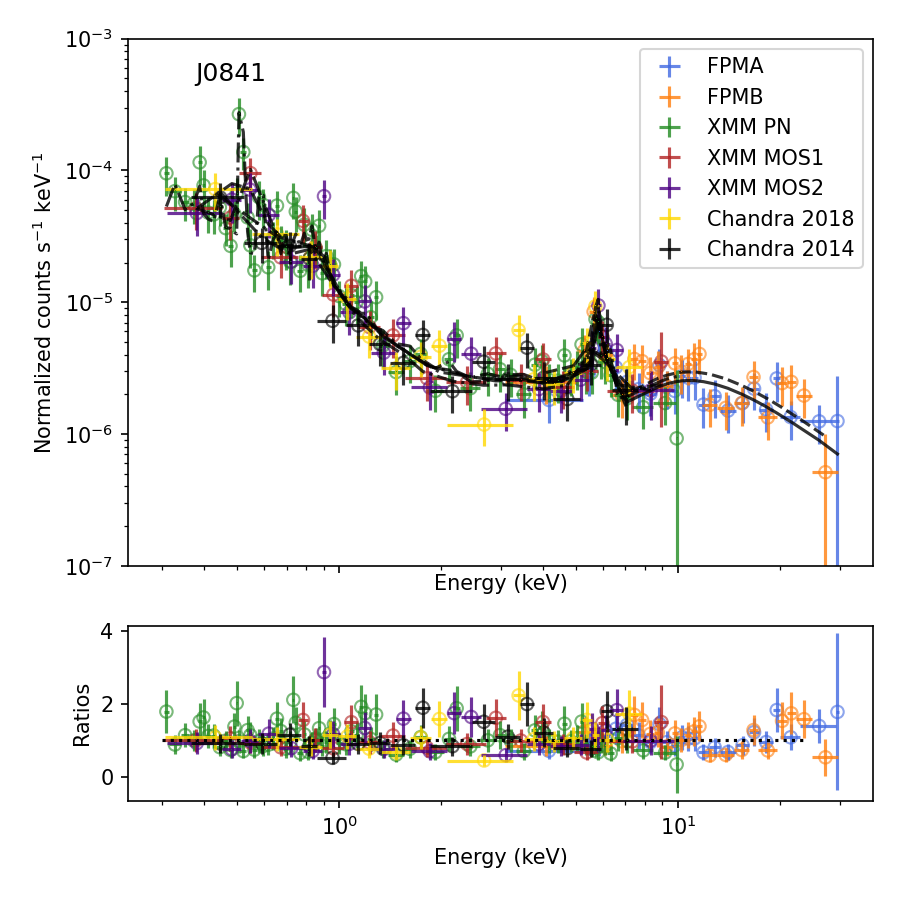}}\\ \vspace{-6mm}
    \subfloat{\includegraphics[width=1.0\linewidth]{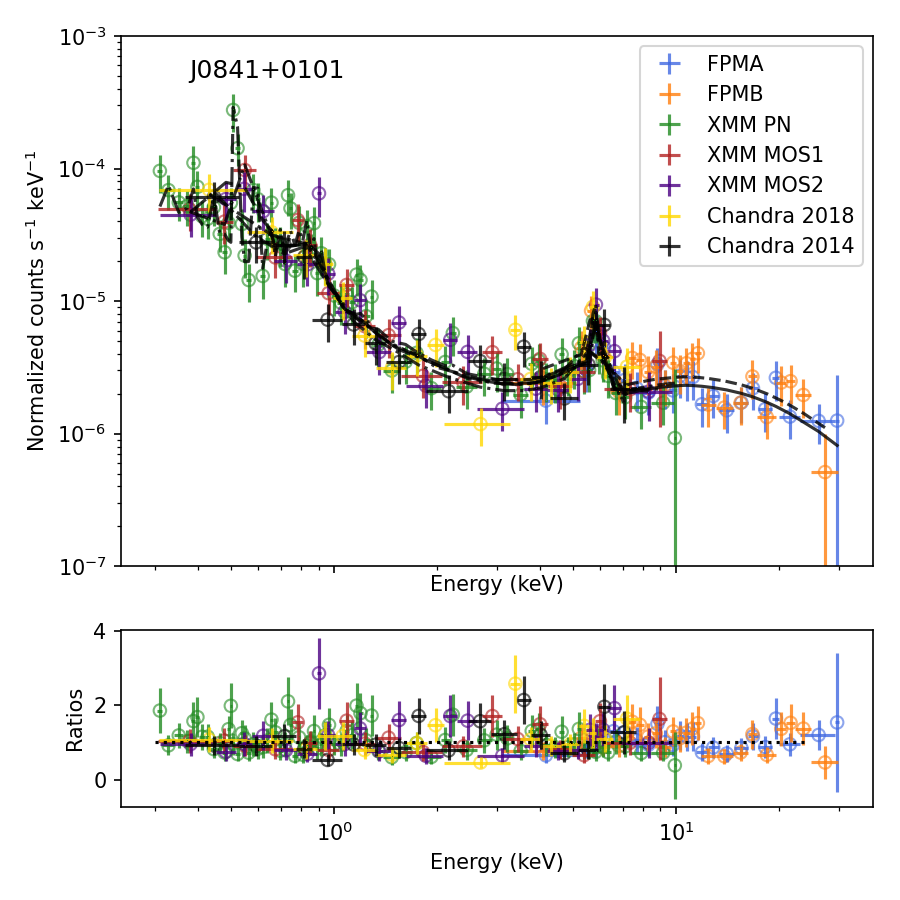}} %\vspace{-6mm}
    \caption{Best fitting models for J0841+0101. (Top): borus model. (Bottom): pexrav model. Each of these best fittin models include two soft thermal components in the 0.1-1\,keV band, components for reflection and scattered emission, and an absorbed power law component that accounts for photoelectric absorption and Compton scattering. }
    \label{fig:my_label}
\end{figure}

\section{Discussion}

\subsection{Heavily Obscured AGNs, or X-ray Underluminous LIRGs?}
\begin{figure}
    \centering
    \includegraphics[width=1.0\linewidth]{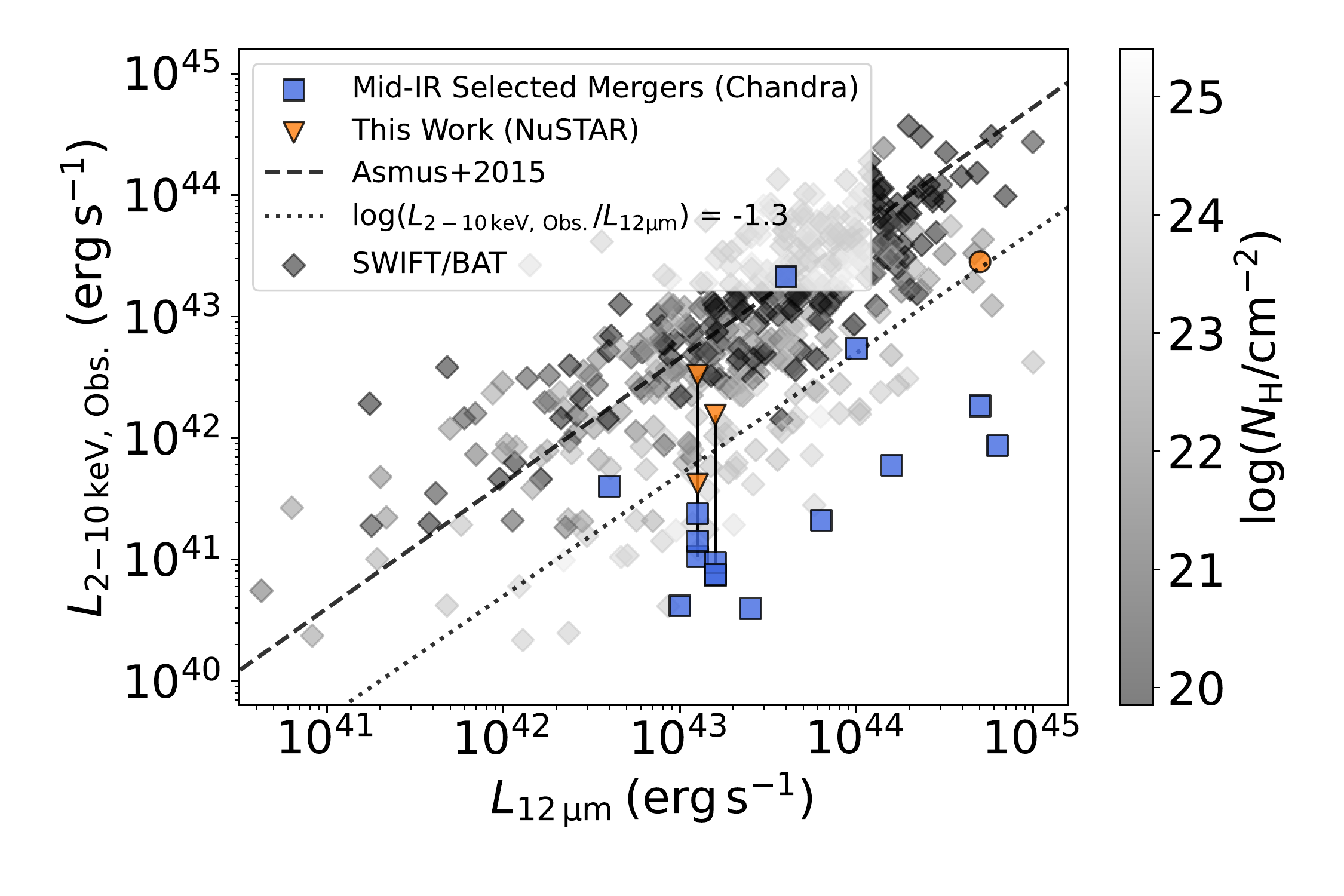}
    \caption{Comparing the intrinsic AGN 12\,$\mu$m luminosity to the total observed $2-10$\,keV luminosity. The intrinsic AGN 12\,$\mu$m luminosity (derived via SED fitting) is given on the abscissa and the total observed $2-10$\,keV luminosity is given on the ordinate. Mid-IR selected mergers from \citet{satyapal2017} and \citet[][]{pfeifle2019a} are given as blue squares; mergers examined in this work with NuSTAR - and hence have 2-10 keV luminosities derived from the NuSTAR - are given in orange, where inverted triangles indicate upper limits and the circle indicates J0841+0101, the only detected source by NuSTAR. Vertical lines connect sources observed with Chandra and NuSTAR. For comparison, we underlay the observed 2-10 keV X-ray and intrinsic AGN 12\,$\mu$m luminosities from the Swift/BAT sample drawn from \citet[][]{ricci2017bass,ichikawa2017,ichikawa2019}; the column densities of the Swift/BAT sample are displayed through the auxiliary color map. The dashed line indicates the expected relation for unobscured AGNs from \citet{asmus2015}, while the dotted line indicates a ratio at which we expect the AGNs to be heavily obscured \citep{pfeifle2022}.}
    \label{fig:lx_v_l12}
\end{figure}

Though NuSTAR could only provide constraints on the nuclear column density in one of the four dual AGN candidates examined here, mid-IR selected dual AGNs are nonetheless expected to be heavily obscured \citep[e.g.,][]{blecha2018}, and this has been observed here for J0841+0101 ($N_{\rm{H}}\approx80-100\times10^{22}$\,cm$^{-2}$) as well as for other AGNs from our mid-IR selected sample: J0849+1114 in \citet{pfeifle2019b}, Mrk 463 \citep{bianchi2008,yamada2018,pfeifle2019a}, NGC 4922 \citep{ricci2017MNRAS,pfeifle2019a}, and J0859+1310 \citep{pfeifle2019a}. If J0122+0100, J1221+1137, J1306+0735, and the other remaining mergers from our mid-IR sample \citep[e.g.,][]{satyapal2017,pfeifle2019a} follow this trend, we should expect nearly all of these systems to host heavily obscured AGNs, as was inferred from the comparison between their observed X-ray luminosities and WISE 12\,$\mu$m luminosities \citep[see Figure 8 from][]{pfeifle2019a}, and shown here in Figure~\ref{fig:lx_v_l12}. While the $12$\,$\mu$m luminosities differ from those shown in our previous work\footnote{Note that the primary difference between Figure~\ref{fig:lx_v_l12} and Figure~8 in \citet{pfeifle2019a} is that the 12\,$\mu$m luminosities were calculated by interpolating between the WISE W2 and W3 luminosities, whereas here they are derived directly from the fits to the SED (see Appendix~\ref{sec:seds}), and thus the $L_{12\,\mu m}$ values differ accordingly.}, Figure~\ref{fig:lx_v_l12} once again shows that the majority of our sample of mid-IR selected dual AGN candidates display substantial X-ray deficits relative to their 12\,$\mu$m luminosities, suggesting the presence of large column densities in these systems. The upper limits derived for J0122+0100, J1221+1137, and J1306+0735 clearly exceed those derived from Chandra; while the Chandra flux levels indicate the AGNs in these system are likely to be obscured, they are likely not as obscured as indicated by the NuSTAR flux upper limits in this work. 

The observed X-ray deficits relative to the mid-IR luminosities observed for J0122+0100, J1221+1137, and J1306+0735 (and other mergers in our sample) may not come as a surprise, given the fact that these mergers are all classified as LIRGs based on their infrared luminosities, i.e., log($L_{\rm{8-1000\,\mu m}}/L_{\odot})>11$; LIRGs are in fact known to be underluminous in X-rays relative to the infrared luminosity \citep{iwasawa2011,torres-alba2018}, regardless of whether they host an AGN. Obscuration is often implicated as the source for the observed X-ray deficits in LIRGs that host AGNs; the presence of heavily obscured AGNs in these mergers is therefore consistent with the general picture that LIRGs are associated with high concentrations of gas and dust and that AGNs identified in late-stage LIRG mergers are often heavily obscured \citep{ricci2017MNRAS,ricci2021}, some of which have also lacked detections in hard X-ray NuSTAR imaging \citep{ricci2017MNRAS,ricci2021}. However, the characteristic LIRG X-ray deficit relative to the infrared is observed both in LIRGs that host AGNs as well as LIRGs that lack AGNs; this point may call into question again whether a population of X-ray binaries could be responsible for the X-ray emission rather than obscured AGNs in these mergers. As a simple test, we compared the X-ray and far-infrared luminosities of these sources to the relations presented in \citet{ranalli2003} and \citet{torres-alba2018}, where the far-infrared luminosity is expected to accurately trace the star formation rate; we calculated the far-infrared luminosities for these three mergers using the prescription outlined in Section~4.5 of \citet{torres-alba2018} after retrieving the 60\,$\mu$m and 100\,$\mu$m flux densities from the IRAS Faint Source Catalog v2.0 \citep{moshir1992}. We found that the X-ray and far-infrared luminosities of these mergers agree with the relation established by \citet[][where Equations 4 and 5 therein were derived for LIRGs that do not show evidence for an AGN in the X-rays or mid-IR;]{torres-alba2018}, and thus the observed X-ray to far-infrared luminosity ratios of these mergers are consistent with those of LIRGs in C-GOALs which do not host AGNs. This is a puzzling result; it would at face value suggest that AGNs may not be necessary to explain the X-ray emission in these mergers, but we cannot use that as evidence to rule out the presence of AGNs. It could be a selection effect: 12\% (18\%) and 10\% (14\%) of C-GOALS (C-GOALS II) objects are non-mergers or pre-mergers while another 30\% (31\%) are early stage mergers \citep{torres-alba2018}; it could be possible that these relations actually vary as a function of merger stage, so the comparison between the late-stage merger LIRGs in this work and the relations derived in \citet{torres-alba2018} may not be straightforward. Similarly, \citet{torres-alba2018} did not differentiate between LIRGs (log[$L_{8-1000\,\rm{\mu m}}/L_{\odot}]>11$) and ULIRGs (log[$L_{8-1000\,\rm{\mu m}}/L_{\odot}]>12$) when deriving these relations, yet the merger fractions are starkly different when comparing LIRGs and ULIRGs \citep[][log$(L_{8-1000\,\rm{\mu m}}$/$L_{\odot})>11.4$ is a rough cutoff threshold for mergers dominating LIRGs.]{kim2013}. Furthermore, this alternative hypothesis is difficult to reconcile with our previous work in \citet{satyapal2017} and \citet{pfeifle2019a} as well as commonplace mid-IR AGN selection criterion \citep{stern2012} and X-ray selection criterion from \citet{iwasawa2011}.  

In \citet{satyapal2017} and \citet{pfeifle2019a}, we used near-IR spectroscopic observations from the Large Binocular Telescope Observatory to estimate the star formation rates \citep[following ][]{kennicutt1994} and the expected X-ray contribution from XRBs \citep[using the relation from ][which relates the star formation rate, stellar mass, and X-ray emission for a galaxy, and was derived from a sample of local LIRGs]{lehmer2010}. In these calculations, we assumed that all of the Pa$\alpha$ emission is due to star formation alone, though in reality an AGN could contribute to this emission as well; in this way, we placed an upper limit on the expected star formation rates and expected XRB X-ray contributions. Comparing the observed X-ray luminosities from Chandra to the expected XRB-driven X-ray emission, we found that our sources are too X-ray luminous to originate from XRBs alone and require an additional source of X-ray emission, i.e., an AGN.  

\citet{iwasawa2011} and \citet{torres-alba2018} adopted an X-ray hardness ratio (HR) selection criterion of HR$>-0.3$ to select X-ray AGNs within C-GOALS I and II. At least one X-ray source in both J0122+0100 and J1306+0735 meet this X-ray selection criterion based upon the reported HRs in \citet{pfeifle2019a}; an additional AGN in both of these mergers just barely miss this cut off (HR$=-0.32$ and HR$=-0.35$). Thus, we have evidence based on X-ray color alone for at least one AGN in each of these mergers as well as evidence that the X-ray emission cannot be explained solely via star formation based on the near-IR. There were other diagnostics as well that we explored previously, such as the detection of a high excitation [Si VI] coronal line in both J0122+0100 and J1221+1137, which we take as an unambiguous signature of AGNs. Similarly, J0122+0100 satisfies the WISE AGN color cut of $W1-W2>0.8$ from \citet{stern2012} and displays similar AGN colors to other bonafide AGNs within our sample.

In what other ways do these objects compare to the C-GOALS LIRG population? Here we specifically compare against C-GOALS II \citep{torres-alba2018} rather than C-GOALS I \citep{iwasawa2011}, as the X-ray luminosities in C-GOALS II more closely match those in our sample here. In addition to the X-ray-to-far-infrared luminosity correlation, we can also examine the X-ray to infrared luminosity ratio, log($L_{\rm{2-10\,keV}}/L_{8-1000\,\rm{\mu m}}$); when considering the total infrared luminosity of the merger system and the combined 2-10\,keV X-ray luminosities reported in \citet{pfeifle2019a}, J0122+0100, J1221+1137, and J1306+0735 have log($L_{\rm{2-10\,keV}}/L_{8-1000\,\rm{\mu m}}$) of $-$3.9, $-$4.3, and $-$3.8. The ratios for J0122+0100 and J1306+0735 suggest that these systems are overluminous (in the $2-10$\,keV band) relative to the LIRGs in C-GOALS II, including the LIRGS that contain AGNs. However, \citet{iwasawa2009} noted that the X-ray to infrared correlation becomes less clear when systems comprising multiple sources are combined and plotted as a single object, so we instead compute distinct log($L_{\rm{2-10\,keV}}/L_{8-1000\,\rm{\mu m}}$) ratios for each nucleus; we assume the following when assigning fractional contributions to the infrared luminosity: (1) the maximum X-ray luminosities attributable to XRBs in the nuclei of these mergers \citep[reported in Table~7 of][]{pfeifle2019a} were derived using near-IR observations and should therefore serve as a proxy for the relative contributions of the nuclei to the near-IR emission, and (2) we can extrapolate this fractional contribution to the total near-IR emission further to the fractional contribution to the total infrared luminosity (an admittedly simplistic assumption). Following this prescription, we find log($L_{\rm{2-10\,keV}}/L_{8-1000\,\rm{\mu m}}$) ratios of $-4.0$ and $-3.6$ in J0122+0100, $-4.2$ in $-4.8$ in J1221+1137, and $-3.5$, $-3.9$, $-3.8$ for the three sources in J1306+0735. Again as before, we find that the sources in J0122+0100 and J1306+0735 are either consistent with, or overluminous relative to, the LIRG population in C-GOALS II when AGNs are included in the LIRG sample (log[$L_{\rm{2-10\,keV}}/L_{8-1000\,\rm{\mu m}}]=-4.04\pm0.48$; see Table~7 in \citealp{torres-alba2018}); these two mergers host X-ray sources that exhibit emission ratios unlike the LIRGs that lack AGNs (log[$L_{\rm{2-10\,keV}}/L_{8-1000\,\rm{\mu m}}]=-4.18\pm0.37$; see Table~7 in \citealp{torres-alba2018}), though they are mostly within the error bounds of the values reported for these ratios in \citet{torres-alba2018} due to the scatter in this correlation. The X-ray to infrared ratios for J1221+1137, on the other hand, are suspiciously similar to LIRGs which lack AGNs; this again is difficult to reconcile with the observation of a high-ionization coronal line in one of the nuclei.

Future works will investigate this issue in more depth for the full sample of mid-IR selected pairs from \citep{satyapal2017} and \citep{pfeifle2019a}, since the majority of the sample showed similar X-ray deficits and low X-ray luminosities. Based upon the available evidence, our preferred interpretation is that these systems contain AGNs that, given the prevalence of gas and dust in LIRGs, are very likely obscured. However, we cannot rule out that star formation contributes a non-negligible fraction to the observed X-ray and infrared luminosities; indeed, obscured star forming regions are also expected in LIRGs, and \citet{pfeifle2019a} found evidence for soft thermal X-ray components in both J0122+0100 and J1221+1137, implying the presence of star formation-driven emission. If it is the case that star formation contributes significantly to the infrared emission in these galaxy mergers, we could be over-predicting the intrinsic X-ray 2-10\,keV luminosities for these sources by using the $L_{\rm{2-10\,keV}}$ vs. $L_{12\rm{\mu m}}$ relation derived by \citet{asmus2015} or other similar relations derived for non-LIRG objects. By extension, then, we would be over-predicting the column densities in these AGNs and AGN candidates, since column densities on the order of $\sim10^{24}$\,cm$^{-2}$ may not be needed to explain the X-ray deficit relative to the mid-IR. Unfortunately, the NuSTAR data cannot differentiate between these scenarios. 

\subsection{The Prevalence of Absorption in Dual AGNs}
One long-standing observation amongst the known population of dual AGNs as well as dual AGN candidates (at all separations, not just limited to $<10$\,kpc) is the prevalence of obscuration along the line-of-sight, where both nuclei are almost ubiquitously found to be obscured by column densities of $N_{\rm{H}}>10^{22}$\,cm$^{-2}$ and at least one (and in some cases both) usually shows higher column densities, on the order of $10^{23}$\,cm$^{-2}$ to $10^{24}$\,cm$^{-2}$. Thus, the expectation that mid-IR selected dual AGNs should be heavily obscured naturally conforms to our present day paradigm. Clear examples of dual AGNs where both nuclei are Compton-thick include NGC 6240 \citep[][]{komossa2003} and SWIFT J2028.5+2543 \citep[][]{koss2016}, while there are several of examples of dual AGNs in which one nucleus is Compton-thick while the other is obscured by $10^{23}$\,cm$^{-2}$ \citep[e.g., IRAS 20210+1121 and Mrk 273,][]{piconcelli2010,iwasawa2018} or $10^{22}$\,cm$^{-2}$ \citep[e.g., Mrk 266][]{iwasawa2020}. Others host nuclei both obscured by $>10^{23}$\,cm$^{-2}$ \citep[e.g., Mrk\,463,][]{bianchi2008} or $>10^{22}$\,cm$^{-2}$ \citep[e.g., ESO 509-IG066,][]{guainazzi2005,kosec2017}. Large samples of optically-selected dual AGNs and (confirmed and candidates) typically reveal a mixture of Compton-thin and Compton-thick levels of obscuration \citep[e.g.,][]{derosa2018,hou2019}, and there are some cases where one nucleus is found to be obscured while the other is unobscured \citep[e.g.,][]{derosa2018,hou2023}. There are, however, puzzling cases where one or both nuclei of dual AGN have been shown to be unobscured, including the Swift-BAT selected Mrk 739 \citep[][though the flat photon indices are suggestive of more obscured AGNs]{koss2011}, J1126+2944 \citep[][though this is a minor merger with mass ratio 460:1, and it is unclear if mergers as minor as these should host heavily obscured AGNs]{comerford2015}, SDSS J1108+0659 and SDSS J1146+5110S \citep[which appear to host unobscured and lightly obscured nuclei,][]{liu2013}. \citet{benitez2022} recently analyzed archival Chandra data for the Swift-BAT selected IRAS 05589+2828 \citep[e.g.,][]{koss2012} and found it consistent with an unobscured Seyfert 1; there were unfortunately too few counts detected from the companion galaxy (2MASX J06021107+2828382) to examine the column density.

There are a large number of dual AGNs that have been confirmed in the literature which lack constraints on the column densities for one of the nuclei, yet these observations do generally still reveal a significant fraction of obscured AGNs, including Was49b \citep[e.g.,][]{secrest2017}, the triple AGN J0849+1114 \citep{pfeifle2019b,liu2019}, SDSS J140737.17+442856.2 \citep[the X-ray bright nucleus shows only $N_{\rm{H}}\sim2\times10^{22}$\,cm$^{-2}$,][but the equivalent width of the Fe K$\alpha$ line, 0.76\,keV, would suggest a larger column density]{ellison2017}, and Arp 299 (constraints are not available for the companion low-luminosity AGN \citep[][]{pereztorres2010}, but \citealp{ptak2015} found that the X-ray bright AGN is Compton-thick).  

While the general consensus has thus far been that dual AGNs generally show high levels of obscuration, this remains to be confirmed using larger and more statistically complete samples of dual AGNs coupled with soft and hard X-ray observations (in concert with sub-millimeter observations, which can also trace molecular gas) to constrain the column densities along the lines-of-sight in these systems. Likewise, possible anti-correlations between the projected pair separation and (1) obscuring column \citep{guainazzi2021} and (2) X-ray 2-10\,keV flux \citep{koss2012} have been reported for dual AGNs, but again require larger and statistically complete samples of dual AGNs to confirm these trends. Indeed, the main limitation to understanding the role of obscuration (and other X-ray properties) in dual AGNs (and candidate) environments is the general lack of X-ray spectroscopic information (or multiwavelength information that provide constraints on obscuring columns) for both nuclei in each case, and hence column density estimates for each nucleus. For now, a comprehensive examination of the known and candidate populations of dual AGNs in terms of their morphological classes, optical types, selection strategies, and column densities may aid in furthering our understanding of the importance and overall prevalence of obscuration in these systems, though that is beyond the scope of this work; the first literature-complete catalog of dual AGNs, binary AGNs, and recoiling AGNs (and all candidate systems) is currently in development and slated for release in Q4 2023, and may be a critically important repository of information for the field of dual AGNs going forward (Pfeifle et al, in prep.)\footnote{If you would like to be involved with the development of this catalog, please reach out to the corresponding author of this paper.}. AGNs in mergers have been shown to possess higher column densities than isolated control AGNs \citep[e.g.,][]{kocevski2015}, but it is not yet clear whether dual AGNs simply fall within this population or if they show even further enhancements of nuclear obscuration than single AGNs in mergers. Understanding whether dual AGNs are generally more obscured than single AGNs in mergers or isolated AGNs may be important to our understanding of the evolution of dual AGN activation and fueling along the merger sequence.

\subsection{Future Prospects for Hard X-ray Dual AGN Science }

Over the last decade \nustar{} has provided access to the hard X-ray band (3-78 keV), but it has still struggled in the area of dual AGN science because: (1) the angular resolution of \nustar{} has so far been fairly prohibitive for identifying distinct X-ray sources in the imaging for all but the nearest dual AGNs and candidates\footnote{But see \citet{ptak2015} and \citet{kosec2017} for examples of \nustar{} observations that could resolve the positions of the two putative AGNs.}, and (2) \nustar{} lacks the sensitivity needed to observe many of these systems in a reasonable amount of time. Figure~\ref{fig:nustarimaging} illustrates these two points: of the four dual AGN candidates observed here with \nustar{}, three remain undetected, and while J0841+0101 was robustly detected and spectroscopically examined, the \nustar{} imaging cannot place constraints on the potential number of sources, and the same can be said of the other three other systems in our sample (NGC\,4922, Mrk\,463, J0849+1114) that have been studied with \nustar{} \citep[][]{ricci2017MNRAS,yamada2018,pfeifle2019b}. It is likely for these reasons, aside from a handful systems observed with \nustar{}  \citep[e.g.,][ and this work]{ptak2015,koss2016,kosec2017,ricci2017MNRAS,iwasawa2020}, that hard X-ray (>10 keV) imaging and spectroscopy is not common in the field. Thus, the future of hard X-ray dual AGN research requires better angular and spectral resolution, and higher sensitivity imaging and spectroscopy.

The High Energy X-ray Probe (HEX-P), a current NASA probe-class mission concept, will feature a pair of High Energy Telescopes (HETs) that would offer broad band (2-80 keV) hard X-ray coverage with a stable PSF of $\sim16$\arcsec{} across the FOV, along with a single Low Energy Telescope (LET) with an energy range of $\sim$0.2-25 keV and stable PSF of $\sim3.5$''. With $\sim40\times$ better sensitivity in the 10-80\,keV band and $12\times$ better angular resolution than NuSTAR, HEX-P could study the X-ray properties of dual AGNs across the merger sequence, from the earliest stage mergers down to even late-stage mergers with nuclear pair separations of $10$\arcsec{} and possibly even down to $5$\arcsec{}. Resolving dual hard X-ray sources in the $10-24$\,keV imaging with separations on the order of $5-10$\arcsec{} would be a huge step above what we can currently achieve with \nustar{} (even if, at the closest-resolved separations, the spectra of the two sources are contaminated by one another) and would offer an incredibly clean and efficient means of confirming dual AGNs. Spatially resolved spectroscopy for slightly larger separations ($\gtrsim10$\arcsec{}) would be another dramatic improvement, enabling constraints on distinct line-of-sight column densities, Fe K$\alpha$ lines, reflection components, etc., in exquisite detail (Pfeifle et al., 2023b, in prep) thanks to HEX-P’s accessibility to the $\gtrsim6-30$ keV (and higher) energy range and high effective area over that broad range. HEX-P will struggle in a similar way to \nustar{} as far as resolving the closest-separation dual AGNs ($<5-6$\arcsec{}) in the hard X-ray imaging, but its sensitivity will be a valuable tool for probing the ‘gross’ X-ray properties of the latest-stage merger systems and, in some cases, may allow one to spectrally deconcolve the two sources, as was done with \nustar{} for Mrk\,463 \citep{yamada2018}.

\subsection{Star Formation-Driven Thermal Components in Dual AGNs}
While obscuration is now a commonly discussed characteristic of dual AGNs and (candidates), a seemingly large fraction of these systems with sufficient counts for spectral analysis also show strong soft X-ray emission which are often well fit via soft X-ray thermal components, as we have found in this work for J0841+0101 and as we found in previous dual AGN candidates \citep[J0122+0100 and J1221+1137][]{pfeifle2019a}. A (non-exhaustive) list of examples from the literature include NGC 6240 \citep{komossa2003}, IRAS 20210+1121 \citep[][]{piconcelli2010}, Mrk 463 \citep[though in this case the soft X-ray emission was modeled using a series of soft X-ray emission lines,][]{bianchi2008}, Arp 299 \citep{ballo2004}, and Mrk 266 \citep{mazzarella2012}, Mrk 739 \citep{inaba2022}, and SDSS J0945+4238 and SDSS J1038+3921 \citep{derosa2018}. Even in cases where soft thermal components are not seen in the X-rays, there is other multiwavelength evidence for starburst activity \citep[like in Mrk 273,][]{iwasawa2018} and generally elevated star formation rates. Soft thermal components are commonly found in AGNs \citep[e.g.,][]{}, so one important question would be: Do dual AGNs show soft X-ray thermal components more frequently than isolated AGNs? This should be generally expected, as it has been shown that star formation rates and AGN fractions are expected to increase as a function of decreasing pair separation \citep[e.g.,][]{ellison2008,ellison2011}, so we should expect that dual AGNs generally show elevated star formation rates and a prevalence of soft X-ray thermal components relative to an isolated control population. A second and related question that we could ask: do dual AGNs show soft X-ray thermal components more frequently than single AGNs in galaxy mergers? As with understanding the prevalence of obscuration in dual AGNs, we are currently limited by the availability of sensitive X-ray observations for large samples of dual AGNs and candidates. Though beyond the scope of this work, future studies focusing on Chandra, XMM-Newton, and eventually eROSITA observations of dual AGNs should investigate whether soft X-ray thermal components, like obscuration, are fairly ubiquitous in the X-ray spectra of dual AGNs and how the occupational fraction of these spectral features compares to general populations of AGNs in mergers or isolated galaxies.

\section{Conclusion}
Dual AGNs in late-stage galaxy mergers are predicted to be heavily obscured by gas and dust \citep{capelo2015,blecha2018}, and indeed many dual AGNs discovered to-date (both in late-stage mergers and at larger separations) have shown high absorbing columns along the line-of-sight, on the order of $>10^{23}-10^{24}$~cm$^{-2}$ \citep[e.g.,][]{komossa2003,bianchi2008,piconcelli2010,mazzarella2012,koss2016,secrest2017,derosa2018,pfeifle2019a,pfeifle2019b}. In this work, we presented new \nustar{} and \xmm{} observations for a subset of the sample of mid-IR selected dual AGNs and dual AGN candidates from \citet{satyapal2017} and \citet{pfeifle2019a}, obtained in an effort to better constrain the column densities along the line-of-sight in these AGNs. We summarize our main conclusions here:
\begin{itemize}
    \item Of the four newly observed mid-IR dual AGN candidates, three were not detected by \nustar{}: J0122+0100, J1221+1137, and J1306+0735. J0841+0101, however, is strongly detected by both \xmm{} and \nustar{}.
    \item The non-detections in J0122+0100, J1221+1137, and J1306+0735 imply there are no hard X-ray emitting AGNs in the $10-24$\,keV band in excess of $1.19\times10^{-13}$, $2.36\times10^{-13}$, and $1.23\times10^{-13}$ erg s$^{-1}$ cm$^{-2}$, respectively, in these systems.  
    \item The upper limits on the fluxes for J0122+0100, J1221+1137, and J1306+0735 imply column densities in excess of log($\nhm{}$)= 24.9, 24.8, and 24.6 (assuming C = 0.99), if we attribute the non-detections to obscuration alone. The $2-10$\,keV flux upper limits from \nustar{} are consistent with the fluxes derived from \chandra{} \citep{pfeifle2019a}. The column density lower limits derived here are also roughly consistent with \citet{pfeifle2019a}, where the differences likely arise due to the difference in column density estimation method.%However, given that the $2-10$\,keV flux upper limits from \nustar{} are higher than the more stringently determined fluxes derived from \chandra{} \citep{pfeifle2019a}, these column density limits are very likely to be overestimates.
    \item While we argue that these AGNs are heavily obscured in order to explain the X-ray deficit relative to the mid-infrared emission in each merger, we cannot rule out the presence of obscured star forming regions that contribute significantly to the infrared luminosity. If this were the case, we would be over-predicting the column densities in these AGNs.
    \item Our broad band spectroscopic analysis of J0841+0101, combining \chandra{}, \xmm{}, and \nustar{} observations, have provided more robust constraints on the X-ray spectral properties of the AGN in this merger. Our best fitting phenomenological model suggests a total line-of-sight column density of \nh{}=$80.1_{-52.8}^{+44.9}\times10^{22}$\,cm$^{-2}$, while our physical torus model suggests a torus obscuring column of log($\nhm{}$)=$23.0_{-0.1}^{+0.1}$ and a line-of-sight column density of \nh{}=$121_{-18}^{+20}\times10^{22}$\,cm$^{-2}$. Our models are also consistent with a power law photon index of $\Gamma = 1.9_{-0.4}^{+0.3}$ or $2.0_{-0.2}^{+0.2}$ (depending on the model), the presence of two thermal components ($T_1=0.17_{-0.06}^{+0.03}$ and $T_2=0.90_{-0.11}^{+0.13}$, or $T_1=0.16_{-0.06}^{+0.04}$ and $T_2=0.90_{-0.12}^{+0.14}$, depending one the model), scattered power law emission ($0.7_{-0.3}^{0.4}$\,\% to $2.7_{-2.2}^{+16.1}$\,\%, depending on the model), and reflection off of an obscuring medium such as a torus.
\end{itemize}
Though the majority of the dual AGN candidates studied here were not detected by \nustar{}, our flux upper limits for J0122+0100, J1221+1137, and J1306+0735, and spectroscopic results for J0841+0101, coupled with the previously published results for Mrk\,463, NGC 4122, and J0849+1114, are consistent with the picture of \wise{}-selected dual AGN systems hosting heavily obscured AGNs. Future hard X-ray analyses of dual AGNs and dual AGN candidates, such as those observed here, will benefit greatly from hard X-ray missions with greater sensitivity and angular resolution, like HEX-P.

\acknowledgments
We thank the anonymous referee for their careful and thoughtful review, which helped to improve this work. We thank George Lansbury for sharing his Python script for calculating Bayesian upper limits. We thank Koji Mukai for helpful discussions on using PIMMS for \nustar{} flux estimations. R.W. P. and S. S. gratefully acknowledge support from NASA grant 80NSSC19K0799. R. W. P. gratefully acknowledges support through an appointment to the NASA Postdoctoral Program at Goddard Space Flight Center, administered by ORAU through a contract with NASA. C. R. acknowledges support from the Fondecyt Iniciacion grant 11190831 and ANID BASAL project FB210003. 

This work makes use of data from the \textit{NuSTAR} mission, a project led by Caltech, managed by the Jet Propulsion Laboratory, and funded by NASA. We thank the \textit{NuSTAR} Operations, Software, and Calibration teams for their support with the execution and analysis of these observations. This research has made use of the \textit{NuSTAR Data} Analysis Software, jointly developed by the ASI Science Data Center (Italy) and Caltech. The scientific results reported in this article are based in part on data obtained from the \textit{Chandra} Data Archive and published previously in cited articles. This research has made use of software provided by the \textit{Chandra} X-ray Center (CXC) in the application packages \textsc{CIAO}. This research has made use of the NASA/IPAC Infrared Science Archive, which is funded by the National Aeronautics and Space Administration and operated by the California Institute of Technology. This publication makes use of data products from the Wide-field Infrared Survey Explorer, which is a joint project of the University of California, Los Angeles, and the Jet Propulsion Laboratory/California Institute of Technology, funded by the National Aeronautics and Space Administration. This publication makes use of data products from the Two Micron All Sky Survey, which is a joint project of the University of Massachusetts and the Infrared Processing and Analysis Center/California Institute of Technology, funded by the National Aeronautics and Space Administration and the National Science Foundation.\\ 

\vspace{5mm}
\facilities{\chandra{}, \xmm{}, \nustar{}, \wise{}, \sdss, Sloan, CTIO:2MASS, FLWO:2MASS, IRAS}

\software{APLpy \citep{robitaille2012}, pandas \citep{mckinney2010}, NumPy \citep{oliphant2006,walt2011,harris2020array}, SciPy \citep{virtanen2020}, \textsc{matplotlib} \citep{hunter2007}, \textsc{heasoft} \citep{heasoft}, \textsc{xspec} \citep{arnaud1996}, \textsc{ciao} \citep{fruscione2006}, \textsc{sas} \citep{2004ASPC..314..759G}, \textsc{nustardas}, \textsc{ds9} \citep{joyce2003}, \textsc{Astropy} \citep{2013A&A...558A..33A,2018AJ....156..123A}}, \textsc{topcat} \citep{2005ASPC..347...29T}

\hfill

\appendix
%\vspace{+2cm}
\section{SED Decomposition} 
%\label{sec:SED Decomposition}
\label{sec:seds}
In order to determine the 8--1000\,\micron\ IR luminosities of our objects (and, by extension, obtain the intrinsic AGN 12\,$\mu$m luminosities for use in Section~\ref{sec:nhest}), we fit their spectral energy distributions (SEDs) using a custom Python code employed in \citet{2018ApJ...858..110P} for the Swift BAT AGNs \citep{2017ApJ...850...74K}. In brief, this code convolves the user's choice of SED templates with the system responses corresponding to their data, and the data are fit via weighted non-negative least-squares, with the weights being the inverse variances of the data. For our data, we combined an AGN template from \citet{2006MNRAS.366..767F}, shown in Figure~1 of \citet{2008MNRAS.386.1252H}, with two templates from \citet{2001ApJ...556..562C} corresponding to the lowest and highest IR luminosity star-forming galaxies, which differ primarily in the equivalent widths of their polycyclic aromatic hydrocarbon (PAH) features and the strength of the IR emission compared with the stellar emission. The AGN template has $W1-W2$/$W2-W3$ synthetic colors of 0.86/2.40, while the low-luminosity and high-luminosity star-forming galaxies have corresponding colors of 0.19/1.72 and 0.82/5.67, so our templates have WISE colors typical of AGNs, spiral galaxies, and LIRGs/ULIRGs \citep[e.g.,][Figure~12]{2010AJ....140.1868W}.

\begin{figure}
\centering
\includegraphics[width=0.99\linewidth]{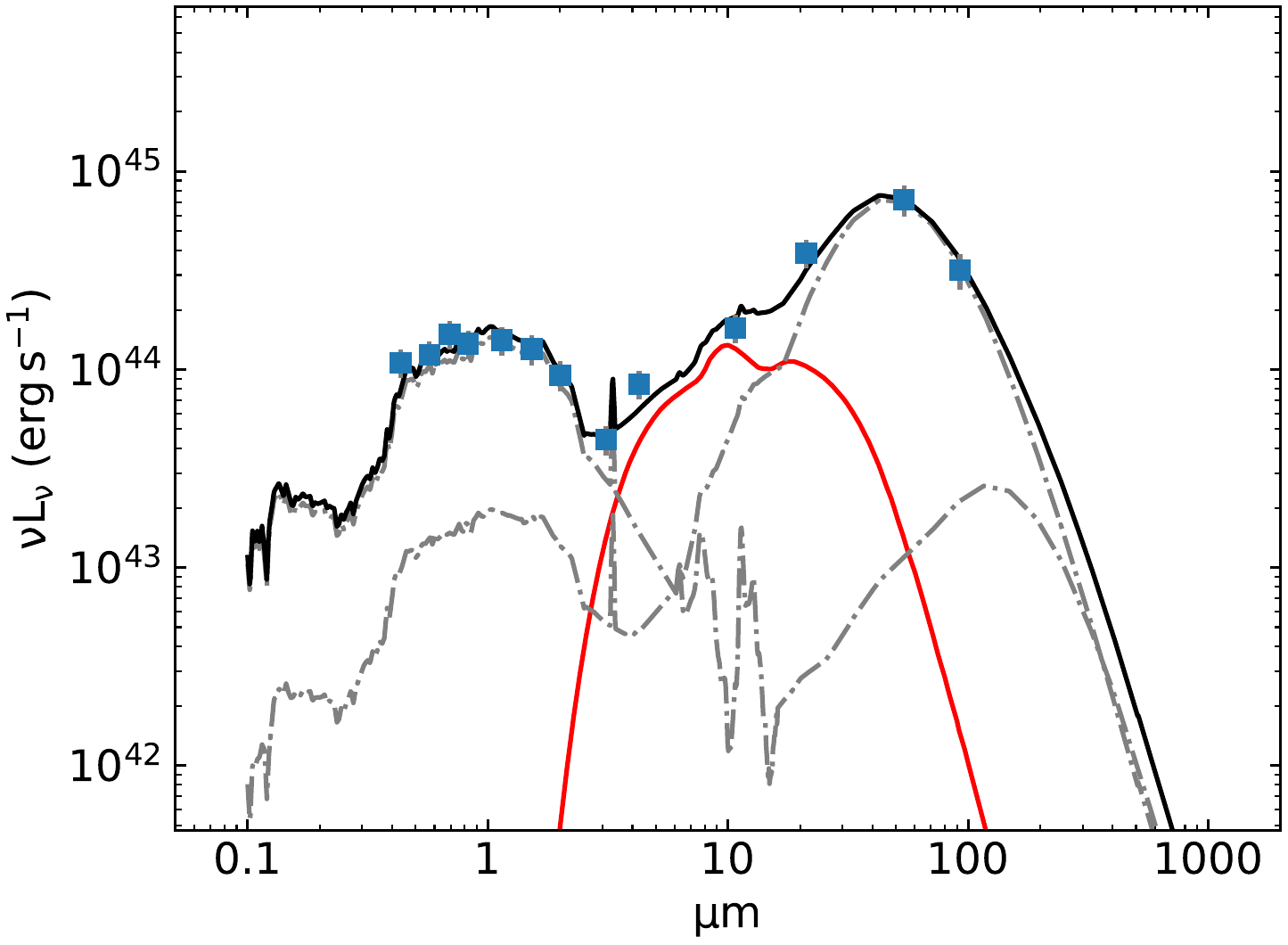}
\caption{Example of one of our SED fits, SDSS J084905.51$+$111447.2. The two dashed-dotted gray subcomponents are the star-forming templates, while the red subcomponent is the reddened AGN. The line-of-sight best-fit extinction of the AGN component is $E(B-V)=16$~mag, indicating that this object is likely Compton-thick with $N_\mathrm{H}\sim1.6\times10^{24}$~cm$^{-2}$. This is within $\sim3\%$ of the value found in \citet{pfeifle2019a} using X-ray data from Chandra and NuSTAR. The wavelength scale is rest-frame.}
\label{fig:J084905.51+111447.2}
\end{figure}

We used photometry from the SDSS~DR12, The Two Micron All Sky Survey \citep[2MASS;][]{2006AJ....131.1163S}, WISE, and the Infrared Astronomical Satellite \citep[IRAS;][]{1984ApJ...278L...1N} appropriate for extended systems. For SDSS, we used the \texttt{modelMag} values, with the exception of the $u$ band, which we exclude due to uncertainties arising from sky level estimates and the known ``red leak''/scattered light issues with the band.\footnote{\url{https://www.sdss.org/dr12/imaging/caveats/}} For 2MASS, we use the Extended Source Catalog (XSC) magnitudes where available, and the Point Source Catalog magnitudes otherwise. We do not use the 2MASS data for SDSS~J130125.26+291849.5, as the relatively large angular extent of this object means that much of its near-IR emission is below the sky brightness limit of the 2MASS survey, so its near-IR emission is underestimated. For WISE, we use the elliptical \texttt{gmag} magnitudes where available, the point-spread-function (PSF)-fit \texttt{mpro} magnitudes otherwise, with the exception of SDSS~J130125.26+291849.5 where we use the large aperture \texttt{mag\_8} magnitudes. For sources with IR flux densities either in the IRAS Point Source Catalog (PSC) or the IRAS Faint Source Catalog (FSC), we additionally use the 60 and 100 \micron\ flux densities, preferentially from the PSC. To convert to the AB system, we added 0.02~mag to the SDSS $z$ band,\footnote{\url{https://www.sdss.org/dr12/algorithms/fluxcal/\#SDSStoAB}} we used the 2MASS Vega/AB offsets available in \textsc{topcat},\footnote{Version 4.6-1; \url{http://www.star.bris.ac.uk/~mbt/topcat}} and we used the standard Vega/AB offsets listed in the WISE documentation.\footnote{\url{http://wise2.ipac.caltech.edu/docs/release/allsky/expsup/sec4_4h.html\#conv2ab}} Finally, we corrected the $g$ through $W2$ magnitudes for Galactic dust extinction using $E(B-V)$ values following \citet{2011ApJ...737..103S}. To account for systematic errors arising due to differences of aperture and flux calibration between the photometry catalogs, as well as differences between the SED templates and the true object SED, we fit each object iteratively, increasing the formal photometric errors in quadrature until the reduced chi-squared of the fit became unity. At each iteration, we fit the AGN component along a grid of $E(B-V)$ values ranging from 0.0 to 50, which corresponds to a neutral hydrogen column density range from zero to $\sim5\times10^{24}$~cm$^{-2}$ \citep{2001A&A...365...28M}. We used the \citet{1998ApJ...500..816G} extinction curve for UV wavelengths and the \citet{1989ApJ...345..245C} extinction curve otherwise. To estimate confidence intervals, we fit each object $10^4$ times, at each iteration drawing a permutation of the magnitudes using the adjusted uncertainties.  We find that our systems have $8-1000~\micron$ luminosities from star formation between $1.2\times10^{10}~L_\sun$ and $5.3\times10^{11}~L_\sun$, with a mean value of $2.7\times10^{11}~L_\sun$, and $80\%$ of our systems are above $10^{11}~L_\sun$, placing them predominantly in the class of luminous infrared galaxies (LIRGs). We show an example of one of our SED fits in Figure~\ref{fig:J084905.51+111447.2} and provide the 8-1000\,$\mu$m and 12\,$\mu$m luminosities in Table~\ref{table:sedfits}.

\begin{deluxetable}{crrr}
\tablecaption{SED Fitting Results}
\label{table:sedfits}
\tablehead{
			 &							   & 									 & 	 \\ [-0.3cm]
\colhead{SDSS} & \colhead{$E^\mathrm{AGN}_{B-V}$} & \colhead{$\log{\left(\frac{L^\mathrm{AGN}_{12 \mu\mathrm{m}}}{\mathrm{erg\,s^{-1}}}\right)}$}  & \colhead{$\log{\left(\frac{L_{8-1000\mu\mathrm{m}}}{L_\sun}\right)}$} \\ [-0.4cm]
}
\startdata
J0122$+$0100 & $0.0^{+11.0}_{-~0.0}$ & $43.1^{+0.7}_{-0.1}$ & $11.3^{+0.1}_{-0.3}$\\
J0841$+$0101 & $15.0^{+~4.0}_{-13.1}$ & $44.7^{+0.2}_{-0.8}$ & $11.7^{+0.3}_{-0.6}$\\
J0849$+$1114 & $16.0^{+~5.0}_{-~5.0}$ & $44.2^{+0.2}_{-0.2}$ & $11.4^{+0.1}_{-0.1}$\\
J0859$+$1310 & $0.7^{+~7.6}_{-~0.3}$ & $42.6^{+0.3}_{-0.2}$ & $10.2^{+0.1}_{-0.2}$\\
J0905$+$3747 & $1.9^{+~5.0}_{-~1.9}$ & $43.1^{+0.3}_{-0.3}$ & $11.2^{+0.1}_{-0.2}$\\
J1036$+$0221 & $11.0^{+~7.0}_{-11.0}$ & $43.8^{+0.3}_{-0.7}$ & $11.7^{+0.1}_{-0.2}$\\
J1045$+$3519 & $0.0^{+11.0}_{-~0.0}$ & $43.2^{+0.6}_{-0.2}$ & $11.6^{+0.1}_{-0.2}$\\
J1126$+$1913 & $7.5^{+11.5}_{-~7.5}$ & $43.4^{+0.5}_{-0.8}$ & $11.5^{+0.1}_{-0.3}$\\
J1147$+$0945 & $0.5^{+~0.2}_{-~0.3}$ & $43.6^{+0.1}_{-0.1}$ & $10.6^{+0.1}_{-0.3}$\\
J1159$+$5320 & $8.2^{+~1.5}_{-~1.6}$ & $43.0^{+0.0}_{-0.1}$ & $10.1^{+0.0}_{-0.1}$\\
J1221$+$1137 & $0.0^{+12.0}_{-~0.0}$ & $43.2^{+0.6}_{-0.3}$ & $11.7^{+0.1}_{-0.1}$\\
J1301$+$2918 & $0.7^{+~4.3}_{-~0.1}$ & $43.2^{+0.2}_{-0.1}$ & $11.3^{+0.0}_{-0.1}$\\
J1306$+$0735 & $0.0^{+22.0}_{-~0.0}$ & $43.1^{+1.1}_{-0.1}$ & $11.6^{+0.0}_{-0.2}$\\
J1356$+$1822 & $3.6^{+~1.4}_{-~3.6}$ & $44.8^{+0.1}_{-0.2}$ & $11.4^{+0.1}_{-0.1}$\\
J2356$-$1016 & $0.0^{+~4.5}_{-~0.0}$ & $44.0^{+0.3}_{-0.1}$ & $11.5^{+0.1}_{-0.2}$
\enddata
\tablecomments{$L^\mathrm{AGN}_{12 \mu\mathrm{m}}$ is the intrinsic, rest-frame 12 \micron\ monochromatic luminosity of the AGN component. $L_{8-1000\,\mu\mathrm{m}}$ is the integrated 8--1000\,\micron\ luminosity of the star-forming templates. The confidence intervals are $5\%-95\%$. AGN reddening of $E(B-V) = 1$ approximately corresponds to $N_\mathrm{H} = 10^{23}$~cm$^{-2}$ (Maiolino et al.~2001).}
\end{deluxetable}

\section{Poorer Spectral Fitting Results for J0841+0101}
\label{sec:apppoorfits}
For clarity, we include here in Table~\ref{apptable:specanalysis} the results of the poorer spectral fits described in Section~\ref{sec:specanalysis}.

\begin{table*}
\begin{center}
\caption{Poorer Spectral Fitting Results for J0841+0101}
\label{apptable:specanalysis}
\begin{tabular}{cccccccccccc}
\hline
\hline
\noalign{\smallskip}
\noalign{\smallskip}
Model & C-stat & d.o.f & $\Gamma$ & $N_{\rm{H}, Torus}$ & $N_{\rm{H}, LOS}$ & R & $f_{\rm{S}}$ & T$_1$ & T$_2$  \\%& $L_{\rm{2-10\,keV}}$ & $L_{\rm{3-24\,keV}}$\\
 &  &  &  & (log[cm$^{2}$]) & ($10^{22}$ cm$^{-2}$) &  & (\%) & (keV) & (keV)  \\%& (erg\,s$^{-1}$) & (erg\,s$^{-1}$) \\
(1) & (2) & (3) & (4) & (5) & (6) & (7) & (8) & (9) & (10) \\%& (12) & (13)\\
\noalign{\smallskip}
\noalign{\smallskip}
\hline
\noalign{\smallskip}
Pexrav & $2156.57$ & $1918$ & $1.5_{-0.04}^{+0.04}$ & \dots & NC & $-10.0_{+10}^{+0.2}$ & NC & \dots & \dots   \\ %pexrav_abspl_scpl
Pexrav & $1943.95$ & $1917$ & $1.4_{-1.4}^{+0.01}$ & $\dots$ & NC & $-10.0_{NC}^{+0.2}$ & $\dots$ & $0.21_{-0.02}^{+0.02}$ & \dots  \\ %pexrav+abspl+apec
Pexrav & $1792.82$ & $1915$ & $2.4_{-0.1}^{NC}$ & $\dots$ & $133.0_{-22.9}^{+26.1}$ & $-0.6_{-0.5}^{+0.3}$ & $\dots$ & $0.19_{-0.03}^{+0.02}$ & $1.05_{-0.12}^{+0.13}$   \\ %pexrav_abspl_apec2
Pexrav & $1788.95$ & $1916$ & $2.2_{-0.3}^{NC}$ & $\dots$ & $91.6_{-28.3}^{+27.5}$ & $-0.8_{-0.9}^{+0.4}$ & $1.0_{-0.7}^{+2.1}$ & $0.22_{-0.03}^{+0.02}$ & $\dots$    \\ %pexrav_abspl_scpl_apec
Pexrav & $1759.23$ & $1914$ & $2.0_{-0.3}^{+0.3}$ & $\dots$ & $86.1_{-34.3}^{+33.4}$ & $-1.4_{-2.2}^{+0.8}$ & $1.3_{-0.9}^{+3.5}$ & $0.17_{-0.06}^{+0.03}$ & $0.91_{-0.12}^{+0.13}$    \\ %pexrav_abspl_scpl_apec2
     
\noalign{\smallskip}
\noalign{\smallskip}
\hline
\noalign{\smallskip}
\noalign{\smallskip}
Borus & $2277.74$ & $1916$ & $1.4_{NC}^{+0.0}$ & $22.1_{-0.1}^{+0.1}$ & $\dots$ & $\dots$ & $\dots$ & $0.27_{-0.01}^{+0.02}$ & \dots    \\ %borus_abspl_apec
Borus & $1861.91$ & $1914$ & $1.4_{NC}^{+0.031}$ & $23.4_{-0.1}^{+0.1}$ & $\dots$ & $\dots$ & $\dots$ & $0.22_{-0.01}^{+0.02}$ & $2.00_{-0.04}^{NC}$     \\ %borus_abspl_apec2
Borus & $1878.45$ & $1917$ & $2.2_{-0.03}^{+0.1}$ & $23.8_{-0.05}^{+0.1}$ & $\dots$ & $\dots$ & $2.14_{-0.3}^{+0.4}$ & $\dots$ & $\dots$    \\ %borus_abspl_scpl
Borus & $1788.86$ & $1915$ & $1.5_{NC}^{+0.1}$ & $23.7_{-0.1}^{+0.1}$ & $\dots$ & $\dots$ & $8.75_{-2.8}^{+2.4}$ & $0.22_{-0.02}^{+0.02}$ & $\dots$  \\ %borus_abspl_scpl_apec
Borus & $1755.74$ & $1913$ & $1.4_{-1.4}^{+0.1}$ & $23.6_{-0.1}^{+0.1}$ & $\dots$ & $\dots$ & $9.06_{-2.1}^{+1.8}$ & $0.17_{-0.17}^{+0.03}$ & $0.90_{-0.14}^{+0.13}$   \\ %borus_abspl_scpl_apec2
\noalign{\smallskip}
\hline
\end{tabular}
\end{center}
\tablecomments{J0841+0101 spectral fitting results from other model permutations. Col. 1: Model choice. Col. 2-3: C-stat and degrees of freedom of the model. Col. 4-6: Photon index, torus column density, and line-of-sight column density. Col. 7-8: reflection coefficient, scattered fraction. Col. 9-10: temperatures for the two apec components. NC: value was unphysical and/or not constrained during fitting.}
\end{table*}

\bibliography{references}{}

\begin{thebibliography}{}
\expandafter\ifx\csname natexlab\endcsname\relax\def\natexlab#1{#1}\fi
\providecommand{\url}[1]{\href{#1}{#1}}
\providecommand{\dodoi}[1]{doi:~\href{http://doi.org/#1}{\nolinkurl{#1}}}
\providecommand{\doeprint}[1]{\href{http://ascl.net/#1}{\nolinkurl{http://ascl.net/#1}}}
\providecommand{\doarXiv}[1]{\href{https://arxiv.org/abs/#1}{\nolinkurl{https://arxiv.org/abs/#1}}}

\bibitem[{{Abazajian} {et~al.}(2009){Abazajian}, {Adelman-McCarthy},
  {Ag{\"u}eros}, {Allam}, {Allende Prieto}, {An}, {Anderson}, {Anderson},
  {Annis}, {Bahcall}, \& et~al.}]{abazajian2009}
{Abazajian}, K.~N., {Adelman-McCarthy}, J.~K., {Ag{\"u}eros}, M.~A., {et~al.}
  2009, \apjs, 182, 543, \dodoi{10.1088/0067-0049/182/2/543}

\bibitem[{{Arnaud}(1996)}]{arnaud1996}
{Arnaud}, K.~A. 1996, in Astronomical Society of the Pacific Conference Series,
  Vol. 101, Astronomical Data Analysis Software and Systems V, ed. G.~H.
  {Jacoby} \& J.~{Barnes}, 17

\bibitem[{{Asmus} {et~al.}(2015){Asmus}, {Gandhi}, {H{\"o}nig}, {Smette}, \&
  {Duschl}}]{asmus2015}
{Asmus}, D., {Gandhi}, P., {H{\"o}nig}, S.~F., {Smette}, A., \& {Duschl}, W.~J.
  2015, \mnras, 454, 766, \dodoi{10.1093/mnras/stv1950}

\bibitem[{{Astropy Collaboration} {et~al.}(2013){Astropy Collaboration},
  {Robitaille}, {Tollerud}, {Greenfield}, {Droettboom}, {Bray}, {Aldcroft},
  {Davis}, {Ginsburg}, {Price-Whelan}, {Kerzendorf}, {Conley}, {Crighton},
  {Barbary}, {Muna}, {Ferguson}, {Grollier}, {Parikh}, {Nair}, {Unther},
  {Deil}, {Woillez}, {Conseil}, {Kramer}, {Turner}, {Singer}, {Fox}, {Weaver},
  {Zabalza}, {Edwards}, {Azalee Bostroem}, {Burke}, {Casey}, {Crawford},
  {Dencheva}, {Ely}, {Jenness}, {Labrie}, {Lim}, {Pierfederici}, {Pontzen},
  {Ptak}, {Refsdal}, {Servillat}, \& {Streicher}}]{2013A&A...558A..33A}
{Astropy Collaboration}, {Robitaille}, T.~P., {Tollerud}, E.~J., {et~al.} 2013,
  \aap, 558, A33, \dodoi{10.1051/0004-6361/201322068}

\bibitem[{{Astropy Collaboration} {et~al.}(2018){Astropy Collaboration},
  {Price-Whelan}, {Sip{\H{o}}cz}, {G{\"u}nther}, {Lim}, {Crawford}, {Conseil},
  {Shupe}, {Craig}, {Dencheva}, {Ginsburg}, {VanderPlas}, {Bradley},
  {P{\'e}rez-Su{\'a}rez}, {de Val-Borro}, {Aldcroft}, {Cruz}, {Robitaille},
  {Tollerud}, {Ardelean}, {Babej}, {Bach}, {Bachetti}, {Bakanov}, {Bamford},
  {Barentsen}, {Barmby}, {Baumbach}, {Berry}, {Biscani}, {Boquien}, {Bostroem},
  {Bouma}, {Brammer}, {Bray}, {Breytenbach}, {Buddelmeijer}, {Burke},
  {Calderone}, {Cano Rodr{\'\i}guez}, {Cara}, {Cardoso}, {Cheedella}, {Copin},
  {Corrales}, {Crichton}, {D'Avella}, {Deil}, {Depagne}, {Dietrich}, {Donath},
  {Droettboom}, {Earl}, {Erben}, {Fabbro}, {Ferreira}, {Finethy}, {Fox},
  {Garrison}, {Gibbons}, {Goldstein}, {Gommers}, {Greco}, {Greenfield},
  {Groener}, {Grollier}, {Hagen}, {Hirst}, {Homeier}, {Horton}, {Hosseinzadeh},
  {Hu}, {Hunkeler}, {Ivezi{\'c}}, {Jain}, {Jenness}, {Kanarek}, {Kendrew},
  {Kern}, {Kerzendorf}, {Khvalko}, {King}, {Kirkby}, {Kulkarni}, {Kumar},
  {Lee}, {Lenz}, {Littlefair}, {Ma}, {Macleod}, {Mastropietro}, {McCully},
  {Montagnac}, {Morris}, {Mueller}, {Mumford}, {Muna}, {Murphy}, {Nelson},
  {Nguyen}, {Ninan}, {N{\"o}the}, {Ogaz}, {Oh}, {Parejko}, {Parley}, {Pascual},
  {Patil}, {Patil}, {Plunkett}, {Prochaska}, {Rastogi}, {Reddy Janga},
  {Sabater}, {Sakurikar}, {Seifert}, {Sherbert}, {Sherwood-Taylor}, {Shih},
  {Sick}, {Silbiger}, {Singanamalla}, {Singer}, {Sladen}, {Sooley},
  {Sornarajah}, {Streicher}, {Teuben}, {Thomas}, {Tremblay}, {Turner},
  {Terr{\'o}n}, {van Kerkwijk}, {de la Vega}, {Watkins}, {Weaver}, {Whitmore},
  {Woillez}, {Zabalza}, \& {Astropy Contributors}}]{2018AJ....156..123A}
{Astropy Collaboration}, {Price-Whelan}, A.~M., {Sip{\H{o}}cz}, B.~M., {et~al.}
  2018, \aj, 156, 123, \dodoi{10.3847/1538-3881/aabc4f}

\bibitem[{{Ballo} {et~al.}(2004){Ballo}, {Braito}, {Della Ceca}, {Maraschi},
  {Tavecchio}, \& {Dadina}}]{ballo2004}
{Ballo}, L., {Braito}, V., {Della Ceca}, R., {et~al.} 2004, \apj, 600, 634,
  \dodoi{10.1086/379887}

\bibitem[{{Balokovi{\'c}} {et~al.}(2018){Balokovi{\'c}}, {Brightman},
  {Harrison}, {Comastri}, {Ricci}, {Buchner}, {Gandhi}, {Farrah}, \&
  {Stern}}]{balokovic2018}
{Balokovi{\'c}}, M., {Brightman}, M., {Harrison}, F.~A., {et~al.} 2018, \apj,
  854, 42, \dodoi{10.3847/1538-4357/aaa7eb}

\bibitem[{{Barnes} \& {Hernquist}(1996)}]{barnes1996}
{Barnes}, J.~E., \& {Hernquist}, L. 1996, \apj, 471, 115,
  \dodoi{10.1086/177957}

\bibitem[{{Barnes} \& {Hernquist}(1991)}]{barnes1991}
{Barnes}, J.~E., \& {Hernquist}, L.~E. 1991, \apjl, 370, L65,
  \dodoi{10.1086/185978}

\bibitem[{{Ben{\'\i}tez} {et~al.}(2022){Ben{\'\i}tez},
  {Jim{\'e}nez-Bail{\'o}n}, {Negrete}, {Ruschel-Dutra},
  {Rodr{\'\i}guez-Espinosa}, {Cruz-Gonz{\'a}lez}, {Rodr{\'\i}guez},
  {Chavushyan}, {Marziani}, {Guti{\'e}rrez}, {Gonz{\'a}lez-Martin}, {Jiang}, \&
  {D'Onofrio}}]{benitez2022}
{Ben{\'\i}tez}, E., {Jim{\'e}nez-Bail{\'o}n}, E., {Negrete}, C.~A., {et~al.}
  2022, \mnras, 516, 5270, \dodoi{10.1093/mnras/stac2244}

\bibitem[{{Bianchi} {et~al.}(2008){Bianchi}, {Chiaberge}, {Piconcelli},
  {Guainazzi}, \& {Matt}}]{bianchi2008}
{Bianchi}, S., {Chiaberge}, M., {Piconcelli}, E., {Guainazzi}, M., \& {Matt},
  G. 2008, \mnras, 386, 105, \dodoi{10.1111/j.1365-2966.2008.13078.x}

\bibitem[{{Blecha} {et~al.}(2018){Blecha}, {Snyder}, {Satyapal}, \&
  {Ellison}}]{blecha2018}
{Blecha}, L., {Snyder}, G.~F., {Satyapal}, S., \& {Ellison}, S.~L. 2018,
  \mnras, 478, 3056, \dodoi{10.1093/mnras/sty1274}

\bibitem[{{Brightman} {et~al.}(2014){Brightman}, {Nandra}, {Salvato}, {Hsu},
  {Aird}, \& {Rangel}}]{brightman2014}
{Brightman}, M., {Nandra}, K., {Salvato}, M., {et~al.} 2014, \mnras, 443, 1999,
  \dodoi{10.1093/mnras/stu1175}

\bibitem[{{Capelo} {et~al.}(2017){Capelo}, {Dotti}, {Volonteri}, {Mayer},
  {Bellovary}, \& {Shen}}]{capelo2017}
{Capelo}, P.~R., {Dotti}, M., {Volonteri}, M., {et~al.} 2017, \mnras, 469,
  4437, \dodoi{10.1093/mnras/stx1067}

\bibitem[{{Capelo} {et~al.}(2015){Capelo}, {Volonteri}, {Dotti}, {Bellovary},
  {Mayer}, \& {Governato}}]{capelo2015}
{Capelo}, P.~R., {Volonteri}, M., {Dotti}, M., {et~al.} 2015, \mnras, 447,
  2123, \dodoi{10.1093/mnras/stu2500}

\bibitem[{{Cardelli} {et~al.}(1989){Cardelli}, {Clayton}, \&
  {Mathis}}]{1989ApJ...345..245C}
{Cardelli}, J.~A., {Clayton}, G.~C., \& {Mathis}, J.~S. 1989, \apj, 345, 245,
  \dodoi{10.1086/167900}

\bibitem[{{Cash}(1979)}]{cash1979}
{Cash}, W. 1979, \apj, 228, 939, \dodoi{10.1086/156922}

\bibitem[{{Chary} \& {Elbaz}(2001)}]{2001ApJ...556..562C}
{Chary}, R., \& {Elbaz}, D. 2001, \apj, 556, 562, \dodoi{10.1086/321609}

\bibitem[{{Comerford} {et~al.}(2012){Comerford}, {Gerke}, {Stern}, {Cooper},
  {Weiner}, {Newman}, {Madsen}, \& {Barrows}}]{comerford2012}
{Comerford}, J.~M., {Gerke}, B.~F., {Stern}, D., {et~al.} 2012, \apj, 753, 42,
  \dodoi{10.1088/0004-637X/753/1/42}

\bibitem[{{Comerford} {et~al.}(2015){Comerford}, {Pooley}, {Barrows}, {Greene},
  {Zakamska}, {Madejski}, \& {Cooper}}]{comerford2015}
{Comerford}, J.~M., {Pooley}, D., {Barrows}, R.~S., {et~al.} 2015, \apj, 806,
  219, \dodoi{10.1088/0004-637X/806/2/219}

\bibitem[{{Comerford} {et~al.}(2011){Comerford}, {Pooley}, {Gerke}, \&
  {Madejski}}]{comerford2011}
{Comerford}, J.~M., {Pooley}, D., {Gerke}, B.~F., \& {Madejski}, G.~M. 2011,
  \apjl, 737, L19, \dodoi{10.1088/2041-8205/737/1/L19}

\bibitem[{{Darg} {et~al.}(2010){Darg}, {Kaviraj}, {Lintott}, {Schawinski},
  {Sarzi}, {Bamford}, {Silk}, {Proctor}, {Andreescu}, {Murray}, {Nichol},
  {Raddick}, {Slosar}, {Szalay}, {Thomas}, \& {Vandenberg}}]{darg2010MNRAS}
{Darg}, D.~W., {Kaviraj}, S., {Lintott}, C.~J., {et~al.} 2010, \mnras, 401,
  1043, \dodoi{10.1111/j.1365-2966.2009.15686.x}

\bibitem[{{De Rosa} {et~al.}(2018){De Rosa}, {Vignali}, {Husemann}, {Bianchi},
  {Bogdanovi{\'c}}, {Guainazzi}, {Herrero-Illana}, {Komossa}, {Kun}, {Loiseau},
  {Paragi}, {Perez-Torres}, \& {Piconcelli}}]{derosa2018}
{De Rosa}, A., {Vignali}, C., {Husemann}, B., {et~al.} 2018, \mnras, 480, 1639,
  \dodoi{10.1093/mnras/sty1867}

\bibitem[{{Ellison} {et~al.}(2011){Ellison}, {Patton}, {Mendel}, \&
  {Scudder}}]{ellison2011}
{Ellison}, S.~L., {Patton}, D.~R., {Mendel}, J.~T., \& {Scudder}, J.~M. 2011,
  \mnras, 418, 2043, \dodoi{10.1111/j.1365-2966.2011.19624.x}

\bibitem[{{Ellison} {et~al.}(2008){Ellison}, {Patton}, {Simard}, \&
  {McConnachie}}]{ellison2008}
{Ellison}, S.~L., {Patton}, D.~R., {Simard}, L., \& {McConnachie}, A.~W. 2008,
  \aj, 135, 1877, \dodoi{10.1088/0004-6256/135/5/1877}

\bibitem[{{Ellison} {et~al.}(2017){Ellison}, {Secrest}, {Mendel}, {Satyapal},
  \& {Simard}}]{ellison2017}
{Ellison}, S.~L., {Secrest}, N.~J., {Mendel}, J.~T., {Satyapal}, S., \&
  {Simard}, L. 2017, \mnras, 470, L49, \dodoi{10.1093/mnrasl/slx076}

\bibitem[{{Ferrarese} \& {Merritt}(2000)}]{ferrarese2000}
{Ferrarese}, L., \& {Merritt}, D. 2000, \apjl, 539, L9, \dodoi{10.1086/312838}

\bibitem[{{Foord} {et~al.}(2020){Foord}, {G{\"u}ltekin}, {Nevin}, {Comerford},
  {Hodges-Kluck}, {Barrows}, {Goulding}, \& {Greene}}]{foord2020}
{Foord}, A., {G{\"u}ltekin}, K., {Nevin}, R., {et~al.} 2020, \apj, 892, 29,
  \dodoi{10.3847/1538-4357/ab72fa}

\bibitem[{{Fritz} {et~al.}(2006){Fritz}, {Franceschini}, \&
  {Hatziminaoglou}}]{2006MNRAS.366..767F}
{Fritz}, J., {Franceschini}, A., \& {Hatziminaoglou}, E. 2006, \mnras, 366,
  767, \dodoi{10.1111/j.1365-2966.2006.09866.x}

\bibitem[{{Fruscione} {et~al.}(2006){Fruscione}, {McDowell}, {Allen},
  {Brickhouse}, {Burke}, {Davis}, {Durham}, {Elvis}, {Galle}, {Harris},
  {Huenemoerder}, {Houck}, {Ishibashi}, {Karovska}, {Nicastro}, {Noble},
  {Nowak}, {Primini}, {Siemiginowska}, {Smith}, \& {Wise}}]{fruscione2006}
{Fruscione}, A., {McDowell}, J.~C., {Allen}, G.~E., {et~al.} 2006, in
  \procspie, Vol. 6270, Society of Photo-Optical Instrumentation Engineers
  (SPIE) Conference Series, 62701V, \dodoi{10.1117/12.671760}

\bibitem[{{Gabriel} {et~al.}(2004){Gabriel}, {Denby}, {Fyfe}, {Hoar}, {Ibarra},
  {Ojero}, {Osborne}, {Saxton}, {Lammers}, \& {Vacanti}}]{2004ASPC..314..759G}
{Gabriel}, C., {Denby}, M., {Fyfe}, D.~J., {et~al.} 2004, in Astronomical
  Society of the Pacific Conference Series, Vol. 314, Astronomical Data
  Analysis Software and Systems (ADASS) XIII, ed. F.~{Ochsenbein}, M.~G.
  {Allen}, \& D.~{Egret}, 759

\bibitem[{{Gebhardt} {et~al.}(2000){Gebhardt}, {Bender}, {Bower}, {Dressler},
  {Faber}, {Filippenko}, {Green}, {Grillmair}, {Ho}, {Kormendy}, {Lauer},
  {Magorrian}, {Pinkney}, {Richstone}, \& {Tremaine}}]{gebhardt2000}
{Gebhardt}, K., {Bender}, R., {Bower}, G., {et~al.} 2000, \apjl, 539, L13,
  \dodoi{10.1086/312840}

\bibitem[{{Gordon} \& {Clayton}(1998)}]{1998ApJ...500..816G}
{Gordon}, K.~D., \& {Clayton}, G.~C. 1998, \apj, 500, 816,
  \dodoi{10.1086/305774}

\bibitem[{{Guainazzi} {et~al.}(2005){Guainazzi}, {Piconcelli},
  {Jim{\'e}nez-Bail{\'o}n}, \& {Matt}}]{guainazzi2005}
{Guainazzi}, M., {Piconcelli}, E., {Jim{\'e}nez-Bail{\'o}n}, E., \& {Matt}, G.
  2005, \aap, 429, L9, \dodoi{10.1051/0004-6361:200400104}

\bibitem[{{Guainazzi} {et~al.}(2021){Guainazzi}, {De Rosa}, {Bianchi},
  {Husemann}, {Bogdanovic}, {Komossa}, {Loiseau}, {Paragi}, {P{\'e}rez-Torres},
  {Piconcelli}, \& {Vignali}}]{guainazzi2021}
{Guainazzi}, M., {De Rosa}, A., {Bianchi}, S., {et~al.} 2021, \mnras, 504, 393,
  \dodoi{10.1093/mnras/stab808}

\bibitem[{{G{\"u}ltekin} {et~al.}(2009){G{\"u}ltekin}, {Richstone}, {Gebhardt},
  {Lauer}, {Tremaine}, {Aller}, {Bender}, {Dressler}, {Faber}, {Filippenko},
  {Green}, {Ho}, {Kormendy}, {Magorrian}, {Pinkney}, \&
  {Siopis}}]{gultekin2009}
{G{\"u}ltekin}, K., {Richstone}, D.~O., {Gebhardt}, K., {et~al.} 2009, \apj,
  698, 198, \dodoi{10.1088/0004-637X/698/1/198}

\bibitem[{Harris {et~al.}(2020)Harris, Millman, van~der Walt, Gommers,
  Virtanen, Cournapeau, Wieser, Taylor, Berg, Smith, Kern, Picus, Hoyer, van
  Kerkwijk, Brett, Haldane, del R{\'{i}}o, Wiebe, Peterson,
  G{\'{e}}rard-Marchant, Sheppard, Reddy, Weckesser, Abbasi, Gohlke, \&
  Oliphant}]{harris2020array}
Harris, C.~R., Millman, K.~J., van~der Walt, S.~J., {et~al.} 2020, Nature, 585,
  357, \dodoi{10.1038/s41586-020-2649-2}

\bibitem[{{Harrison} {et~al.}(2013){Harrison}, {Craig}, {Christensen},
  {Hailey}, {Zhang}, {Boggs}, {Stern}, {Cook}, {Forster}, {Giommi},
  {Grefenstette}, {Kim}, {Kitaguchi}, {Koglin}, {Madsen}, {Mao}, {Miyasaka},
  {Mori}, {Perri}, {Pivovaroff}, {Puccetti}, {Rana}, {Westergaard}, {Willis},
  {Zoglauer}, {An}, {Bachetti}, {Barri{\`e}re}, {Bellm}, {Bhalerao},
  {Brejnholt}, {Fuerst}, {Liebe}, {Markwardt}, {Nynka}, {Vogel}, {Walton},
  {Wik}, {Alexander}, {Cominsky}, {Hornschemeier}, {Hornstrup}, {Kaspi},
  {Madejski}, {Matt}, {Molendi}, {Smith}, {Tomsick}, {Ajello}, {Ballantyne},
  {Balokovi{\'c}}, {Barret}, {Bauer}, {Blandford}, {Brandt}, {Brenneman},
  {Chiang}, {Chakrabarty}, {Chenevez}, {Comastri}, {Dufour}, {Elvis}, {Fabian},
  {Farrah}, {Fryer}, {Gotthelf}, {Grindlay}, {Helfand}, {Krivonos}, {Meier},
  {Miller}, {Natalucci}, {Ogle}, {Ofek}, {Ptak}, {Reynolds}, {Rigby},
  {Tagliaferri}, {Thorsett}, {Treister}, \& {Urry}}]{harrison2013}
{Harrison}, F.~A., {Craig}, W.~W., {Christensen}, F.~E., {et~al.} 2013, \apj,
  770, 103, \dodoi{10.1088/0004-637X/770/2/103}

\bibitem[{{Hatziminaoglou} {et~al.}(2008){Hatziminaoglou}, {Fritz},
  {Franceschini}, {Afonso-Luis}, {Hern{\'a}n-Caballero}, {P{\'e}rez-Fournon},
  {Serjeant}, {Lonsdale}, {Oliver}, {Rowan-Robinson}, {Shupe}, {Smith}, \&
  {Surace}}]{2008MNRAS.386.1252H}
{Hatziminaoglou}, E., {Fritz}, J., {Franceschini}, A., {et~al.} 2008, \mnras,
  386, 1252, \dodoi{10.1111/j.1365-2966.2008.13119.x}

\bibitem[{{Hopkins} {et~al.}(2006){Hopkins}, {Hernquist}, {Cox}, {Di Matteo},
  {Robertson}, \& {Springel}}]{hopkins2006}
{Hopkins}, P.~F., {Hernquist}, L., {Cox}, T.~J., {et~al.} 2006, \apjs, 163, 1,
  \dodoi{10.1086/499298}

\bibitem[{{Hopkins} {et~al.}(2008){Hopkins}, {Hernquist}, {Cox}, \&
  {Kere{\v{s}}}}]{hopkins2008}
{Hopkins}, P.~F., {Hernquist}, L., {Cox}, T.~J., \& {Kere{\v{s}}}, D. 2008,
  \apjs, 175, 356, \dodoi{10.1086/524362}

\bibitem[{{Hou} {et~al.}(2023){Hou}, {Li}, {Liu}, {Li}, {Li}, {Wang}, {Wang},
  \& {Ho}}]{hou2023}
{Hou}, M., {Li}, Z., {Liu}, X., {et~al.} 2023, \apj, 943, 50,
  \dodoi{10.3847/1538-4357/acaaf9}

\bibitem[{{Hou} {et~al.}(2019){Hou}, {Liu}, {Guo}, {Li}, {Shen}, \&
  {Green}}]{hou2019}
{Hou}, M., {Liu}, X., {Guo}, H., {et~al.} 2019, \apj, 882, 41,
  \dodoi{10.3847/1538-4357/ab3225}

\bibitem[{Hunter(2007)}]{hunter2007}
Hunter, J.~D. 2007, Computing in Science \& Engineering, 9, 90,
  \dodoi{10.1109/MCSE.2007.55}

\bibitem[{{Ichikawa} {et~al.}(2017){Ichikawa}, {Ricci}, {Ueda}, {Matsuoka},
  {Toba}, {Kawamuro}, {Trakhtenbrot}, \& {Koss}}]{ichikawa2017}
{Ichikawa}, K., {Ricci}, C., {Ueda}, Y., {et~al.} 2017, \apj, 835, 74,
  \dodoi{10.3847/1538-4357/835/1/74}

\bibitem[{{Ichikawa} {et~al.}(2019){Ichikawa}, {Ricci}, {Ueda}, {Bauer},
  {Kawamuro}, {Koss}, {Oh}, {Rosario}, {Shimizu}, {Stalevski}, {Fuller},
  {Packham}, \& {Trakhtenbrot}}]{ichikawa2019}
---. 2019, \apj, 870, 31, \dodoi{10.3847/1538-4357/aaef8f}

\bibitem[{{Inaba} {et~al.}(2022){Inaba}, {Ueda}, {Yamada}, {Ogawa}, {Uematsu},
  {Tanimoto}, \& {Ricci}}]{inaba2022}
{Inaba}, K., {Ueda}, Y., {Yamada}, S., {et~al.} 2022, \apj, 939, 88,
  \dodoi{10.3847/1538-4357/ac97ec}

\bibitem[{{Iwasawa} {et~al.}(2009){Iwasawa}, {Sanders}, {Evans}, {Mazzarella},
  {Armus}, \& {Surace}}]{iwasawa2009}
{Iwasawa}, K., {Sanders}, D.~B., {Evans}, A.~S., {et~al.} 2009, \apjl, 695,
  L103, \dodoi{10.1088/0004-637X/695/1/L103}

\bibitem[{{Iwasawa} {et~al.}(2018){Iwasawa}, {U}, {Mazzarella}, {Medling},
  {Sanders}, \& {Evans}}]{iwasawa2018}
{Iwasawa}, K., {U}, V., {Mazzarella}, J.~M., {et~al.} 2018, \aap, 611, A71,
  \dodoi{10.1051/0004-6361/201731662}

\bibitem[{{Iwasawa} {et~al.}(2011){Iwasawa}, {Sanders}, {Teng}, {U}, {Armus},
  {Evans}, {Howell}, {Komossa}, {Mazzarella}, {Petric}, {Surace}, {Vavilkin},
  {Veilleux}, \& {Trentham}}]{iwasawa2011}
{Iwasawa}, K., {Sanders}, D.~B., {Teng}, S.~H., {et~al.} 2011, \aap, 529, A106,
  \dodoi{10.1051/0004-6361/201015264}

\bibitem[{{Iwasawa} {et~al.}(2020){Iwasawa}, {Ricci}, {Privon},
  {Torres-Alb{\`a}}, {Inami}, {Charmandaris}, {Evans}, {Mazzarella}, \&
  {D{\'\i}az-Santos}}]{iwasawa2020}
{Iwasawa}, K., {Ricci}, C., {Privon}, G.~C., {et~al.} 2020, \aap, 640, A95,
  \dodoi{10.1051/0004-6361/202038513}

\bibitem[{{Joye} \& {Mandel}(2003)}]{joyce2003}
{Joye}, W.~A., \& {Mandel}, E. 2003, in Astronomical Society of the Pacific
  Conference Series, Vol. 295, Astronomical Data Analysis Software and Systems
  XII, ed. H.~E. {Payne}, R.~I. {Jedrzejewski}, \& R.~N. {Hook}, 489

\bibitem[{{Kennicutt} {et~al.}(1994){Kennicutt}, {Tamblyn}, \&
  {Congdon}}]{kennicutt1994}
{Kennicutt}, Robert~C., J., {Tamblyn}, P., \& {Congdon}, C.~E. 1994, \apj, 435,
  22, \dodoi{10.1086/174790}

\bibitem[{{Kim} {et~al.}(2013){Kim}, {Evans}, {Vavilkin}, {Armus},
  {Mazzarella}, {Sheth}, {Surace}, {Haan}, {Howell}, {D{\'\i}az-Santos},
  {Petric}, {Iwasawa}, {Privon}, \& {Sanders}}]{kim2013}
{Kim}, D.~C., {Evans}, A.~S., {Vavilkin}, T., {et~al.} 2013, \apj, 768, 102,
  \dodoi{10.1088/0004-637X/768/2/102}

\bibitem[{{Kocevski} {et~al.}(2015){Kocevski}, {Brightman}, {Nandra},
  {Koekemoer}, {Salvato}, {Aird}, {Bell}, {Hsu}, {Kartaltepe}, {Koo}, {Lotz},
  {McIntosh}, {Mozena}, {Rosario}, \& {Trump}}]{kocevski2015}
{Kocevski}, D.~D., {Brightman}, M., {Nandra}, K., {et~al.} 2015, \apj, 814,
  104, \dodoi{10.1088/0004-637X/814/2/104}

\bibitem[{{Komossa} {et~al.}(2003){Komossa}, {Burwitz}, {Hasinger}, {Predehl},
  {Kaastra}, \& {Ikebe}}]{komossa2003}
{Komossa}, S., {Burwitz}, V., {Hasinger}, G., {et~al.} 2003, \apjl, 582, L15,
  \dodoi{10.1086/346145}

\bibitem[{{Kosec} {et~al.}(2017){Kosec}, {Brightman}, {Stern},
  {M{\"u}ller-S{\'a}nchez}, {Koss}, {Oh}, {Assef}, {Gandhi}, {Harrison}, {Jun},
  {Masini}, {Ricci}, {Walton}, {Treister}, {Comerford}, \&
  {Privon}}]{kosec2017}
{Kosec}, P., {Brightman}, M., {Stern}, D., {et~al.} 2017, \apj, 850, 168,
  \dodoi{10.3847/1538-4357/aa932e}

\bibitem[{{Koss} {et~al.}(2012){Koss}, {Mushotzky}, {Treister}, {Veilleux},
  {Vasudevan}, \& {Trippe}}]{koss2012}
{Koss}, M., {Mushotzky}, R., {Treister}, E., {et~al.} 2012, \apjl, 746, L22,
  \dodoi{10.1088/2041-8205/746/2/L22}

\bibitem[{{Koss} {et~al.}(2011){Koss}, {Mushotzky}, {Treister}, {Veilleux},
  {Vasudevan}, {Miller}, {Sanders}, {Schawinski}, \& {Trippe}}]{koss2011}
---. 2011, \apjl, 735, L42, \dodoi{10.1088/2041-8205/735/2/L42}

\bibitem[{{Koss} {et~al.}(2017){Koss}, {Trakhtenbrot}, {Ricci}, {Lamperti},
  {Oh}, {Berney}, {Schawinski}, {Balokovi{\'c}}, {Baronchelli}, {Crenshaw},
  {Fischer}, {Gehrels}, {Harrison}, {Hashimoto}, {Hogg}, {Ichikawa}, {Masetti},
  {Mushotzky}, {Sartori}, {Stern}, {Treister}, {Ueda}, {Veilleux}, \&
  {Winter}}]{2017ApJ...850...74K}
{Koss}, M., {Trakhtenbrot}, B., {Ricci}, C., {et~al.} 2017, \apj, 850, 74,
  \dodoi{10.3847/1538-4357/aa8ec9}

\bibitem[{{Koss} {et~al.}(2016){Koss}, {Glidden}, {Balokovi{\'c}}, {Stern},
  {Lamperti}, {Assef}, {Bauer}, {Ballantyne}, {Boggs}, {Craig}, {Farrah},
  {F{\"u}rst}, {Gandhi}, {Gehrels}, {Hailey}, {Harrison}, {Markwardt},
  {Masini}, {Ricci}, {Treister}, {Walton}, \& {Zhang}}]{koss2016}
{Koss}, M.~J., {Glidden}, A., {Balokovi{\'c}}, M., {et~al.} 2016, \apjl, 824,
  L4, \dodoi{10.3847/2041-8205/824/1/L4}

\bibitem[{{Koss} {et~al.}(2018){Koss}, {Blecha}, {Bernhard}, {Hung}, {Lu},
  {Trakhtenbrot}, {Treister}, {Weigel}, {Sartori}, {Mushotzky}, {Schawinski},
  {Ricci}, {Veilleux}, \& {Sanders}}]{koss2018}
{Koss}, M.~J., {Blecha}, L., {Bernhard}, P., {et~al.} 2018, \nat, 563, 214,
  \dodoi{10.1038/s41586-018-0652-7}

\bibitem[{{Kraft} {et~al.}(1991){Kraft}, {Burrows}, \& {Nousek}}]{kraft1991}
{Kraft}, R.~P., {Burrows}, D.~N., \& {Nousek}, J.~A. 1991, \apj, 374, 344,
  \dodoi{10.1086/170124}

\bibitem[{{Lansbury} {et~al.}(2014){Lansbury}, {Alexander}, {Del Moro},
  {Gandhi}, {Assef}, {Stern}, {Aird}, {Ballantyne}, {Balokovi{\'c}}, {Bauer},
  {Boggs}, {Brandt}, {Christensen}, {Craig}, {Elvis}, {Grefenstette}, {Hailey},
  {Harrison}, {Hickox}, {Koss}, {LaMassa}, {Luo}, {Mullaney}, {Teng}, {Urry},
  \& {Zhang}}]{lansbury2014}
{Lansbury}, G.~B., {Alexander}, D.~M., {Del Moro}, A., {et~al.} 2014, \apj,
  785, 17, \dodoi{10.1088/0004-637X/785/1/17}

\bibitem[{{Lansbury} {et~al.}(2015){Lansbury}, {Gandhi}, {Alexander}, {Assef},
  {Aird}, {Annuar}, {Ballantyne}, {Balokovi{\'c}}, {Bauer}, {Boggs}, {Brandt},
  {Brightman}, {Christensen}, {Civano}, {Comastri}, {Craig}, {Del Moro},
  {Grefenstette}, {Hailey}, {Harrison}, {Hickox}, {Koss}, {LaMassa}, {Luo},
  {Puccetti}, {Stern}, {Treister}, {Vignali}, {Zappacosta}, \&
  {Zhang}}]{lansbury2015}
{Lansbury}, G.~B., {Gandhi}, P., {Alexander}, D.~M., {et~al.} 2015, \apj, 809,
  115, \dodoi{10.1088/0004-637X/809/2/115}

\bibitem[{{Lansbury} {et~al.}(2017){Lansbury}, {Stern}, {Aird}, {Alexander},
  {Fuentes}, {Harrison}, {Treister}, {Bauer}, {Tomsick}, {Balokovi{\'c}}, {Del
  Moro}, {Gandhi}, {Ajello}, {Annuar}, {Ballantyne}, {Boggs}, {Brandt},
  {Brightman}, {Chen}, {Christensen}, {Civano}, {Comastri}, {Craig}, {Forster},
  {Grefenstette}, {Hailey}, {Hickox}, {Jiang}, {Jun}, {Koss}, {Marchesi},
  {Melo}, {Mullaney}, {Noirot}, {Schulze}, {Walton}, {Zappacosta}, \&
  {Zhang}}]{lansbury2017}
{Lansbury}, G.~B., {Stern}, D., {Aird}, J., {et~al.} 2017, \apj, 836, 99,
  \dodoi{10.3847/1538-4357/836/1/99}

\bibitem[{{Lanzuisi} {et~al.}(2018){Lanzuisi}, {Civano}, {Marchesi},
  {Comastri}, {Brusa}, {Gilli}, {Vignali}, {Zamorani}, {Brightman},
  {Griffiths}, \& {Koekemoer}}]{lanzuisi2018}
{Lanzuisi}, G., {Civano}, F., {Marchesi}, S., {et~al.} 2018, \mnras, 480, 2578,
  \dodoi{10.1093/mnras/sty2025}

\bibitem[{{Lehmer} {et~al.}(2010){Lehmer}, {Alexander}, {Bauer}, {Brandt},
  {Goulding}, {Jenkins}, {Ptak}, \& {Roberts}}]{lehmer2010}
{Lehmer}, B.~D., {Alexander}, D.~M., {Bauer}, F.~E., {et~al.} 2010, \apj, 724,
  559, \dodoi{10.1088/0004-637X/724/1/559}

\bibitem[{{Lintott} {et~al.}(2008){Lintott}, {Schawinski}, {Slosar}, {Land},
  {Bamford}, {Thomas}, {Raddick}, {Nichol}, {Szalay}, {Andreescu}, {Murray}, \&
  {Vandenberg}}]{lintott2008}
{Lintott}, C.~J., {Schawinski}, K., {Slosar}, A., {et~al.} 2008, \mnras, 389,
  1179, \dodoi{10.1111/j.1365-2966.2008.13689.x}

\bibitem[{{Liu} {et~al.}(2013){Liu}, {Civano}, {Shen}, {Green}, {Greene}, \&
  {Strauss}}]{liu2013}
{Liu}, X., {Civano}, F., {Shen}, Y., {et~al.} 2013, \apj, 762, 110,
  \dodoi{10.1088/0004-637X/762/2/110}

\bibitem[{{Liu} {et~al.}(2010){Liu}, {Shen}, {Strauss}, \& {Greene}}]{liu2010}
{Liu}, X., {Shen}, Y., {Strauss}, M.~A., \& {Greene}, J.~E. 2010, \apj, 708,
  427, \dodoi{10.1088/0004-637X/708/1/427}

\bibitem[{{Liu} {et~al.}(2011){Liu}, {Shen}, {Strauss}, \& {Hao}}]{liu2011}
{Liu}, X., {Shen}, Y., {Strauss}, M.~A., \& {Hao}, L. 2011, \apj, 737, 101,
  \dodoi{10.1088/0004-637X/737/2/101}

\bibitem[{{Liu} {et~al.}(2019){Liu}, {Hou}, {Li}, {Nyland}, {Guo}, {Kong},
  {Shen}, {Wrobel}, \& {Peng}}]{liu2019}
{Liu}, X., {Hou}, M., {Li}, Z., {et~al.} 2019, \apj, 887, 90,
  \dodoi{10.3847/1538-4357/ab54c3}

\bibitem[{{Maiolino} {et~al.}(2001){Maiolino}, {Marconi}, {Salvati},
  {Risaliti}, {Severgnini}, {Oliva}, {La Franca}, \&
  {Vanzi}}]{2001A&A...365...28M}
{Maiolino}, R., {Marconi}, A., {Salvati}, M., {et~al.} 2001, \aap, 365, 28,
  \dodoi{10.1051/0004-6361:20000177}

\bibitem[{{Marchesi} {et~al.}(2018){Marchesi}, {Ajello}, {Marcotulli},
  {Comastri}, {Lanzuisi}, \& {Vignali}}]{marchesi2018}
{Marchesi}, S., {Ajello}, M., {Marcotulli}, L., {et~al.} 2018, \apj, 854, 49,
  \dodoi{10.3847/1538-4357/aaa410}

\bibitem[{{Mazzarella} {et~al.}(2012){Mazzarella}, {Iwasawa}, {Vavilkin},
  {Armus}, {Kim}, {Bothun}, {Evans}, {Spoon}, {Haan}, {Howell}, {Lord},
  {Marshall}, {Ishida}, {Xu}, {Petric}, {Sanders}, {Surace}, {Appleton},
  {Chan}, {Frayer}, {Inami}, {Khachikian}, {Madore}, {Privon}, {Sturm}, {U}, \&
  {Veilleux}}]{mazzarella2012}
{Mazzarella}, J.~M., {Iwasawa}, K., {Vavilkin}, T., {et~al.} 2012, \aj, 144,
  125, \dodoi{10.1088/0004-6256/144/5/125}

\bibitem[{McKinney(2010)}]{mckinney2010}
McKinney, W. 2010, in Proceedings of the 9th Python in Science Conference, ed.
  S.~van~der Walt \& J.~Millman, 51 -- 56

\bibitem[{{Mihos} \& {Hernquist}(1996)}]{mihos1996}
{Mihos}, J.~C., \& {Hernquist}, L. 1996, \apj, 464, 641, \dodoi{10.1086/177353}

\bibitem[{{Moshir} {et~al.}(1992){Moshir}, {Kopman}, \& {Conrow}}]{moshir1992}
{Moshir}, M., {Kopman}, G., \& {Conrow}, T.~A.~O. 1992, {IRAS Faint Source
  Survey, Explanatory supplement version 2}

\bibitem[{{M{\"u}ller-S{\'a}nchez} {et~al.}(2015){M{\"u}ller-S{\'a}nchez},
  {Comerford}, {Nevin}, {Barrows}, {Cooper}, \& {Greene}}]{mullersanchez2015}
{M{\"u}ller-S{\'a}nchez}, F., {Comerford}, J.~M., {Nevin}, R., {et~al.} 2015,
  \apj, 813, 103, \dodoi{10.1088/0004-637X/813/2/103}

\bibitem[{{Mushotzky} {et~al.}(1993){Mushotzky}, {Done}, \&
  {Pounds}}]{mushotzky1993}
{Mushotzky}, R.~F., {Done}, C., \& {Pounds}, K.~A. 1993, \araa, 31, 717,
  \dodoi{10.1146/annurev.aa.31.090193.003441}

\bibitem[{{Nasa High Energy Astrophysics Science Archive Research
  Center}(2014)}]{heasoft}
{Nasa High Energy Astrophysics Science Archive Research Center}. 2014,
  {HEAsoft: Unified Release of FTOOLS and XANADU}, Astrophysics Source Code
  Library, record ascl:1408.004.
\newblock \doeprint{1408.004}

\bibitem[{{Neugebauer} {et~al.}(1984){Neugebauer}, {Habing}, {van Duinen},
  {Aumann}, {Baud}, {Beichman}, {Beintema}, {Boggess}, {Clegg}, {de Jong},
  {Emerson}, {Gautier}, {Gillett}, {Harris}, {Hauser}, {Houck}, {Jennings},
  {Low}, {Marsden}, {Miley}, {Olnon}, {Pottasch}, {Raimond}, {Rowan-Robinson},
  {Soifer}, {Walker}, {Wesselius}, \& {Young}}]{1984ApJ...278L...1N}
{Neugebauer}, G., {Habing}, H.~J., {van Duinen}, R., {et~al.} 1984, \apjl, 278,
  L1, \dodoi{10.1086/184209}

\bibitem[{Oliphant(2006)}]{oliphant2006}
Oliphant, T.~E. 2006, A guide to NumPy, Vol.~1 (Trelgol Publishing USA)

\bibitem[{{P{\'e}rez-Torres} {et~al.}(2010){P{\'e}rez-Torres}, {Alberdi},
  {Romero-Ca{\~n}izales}, \& {Bondi}}]{pereztorres2010}
{P{\'e}rez-Torres}, M.~A., {Alberdi}, A., {Romero-Ca{\~n}izales}, C., \&
  {Bondi}, M. 2010, \aap, 519, L5, \dodoi{10.1051/0004-6361/201015462}

\bibitem[{{Perri} {et~al.}(2021){Perri}, {Puccetti}, {Spagnuolo}, {Ficcadenti},
  {Severo}, {Davis}, {Forster}, {Grefenstette}, {Harrison}, \&
  {Madsen}}]{nustardas}
{Perri}, M., {Puccetti}, S., {Spagnuolo}, N., {et~al.} 2021, The NuSTAR Data
  Analysis Software Guide, Version 1.9.7, ASI Space Science Data Center and
  California Institute of Technology.
\newblock
  \url{https://heasarc.gsfc.nasa.gov/docs/nustar/analysis/nustar_swguide.pdf}

\bibitem[{{Pfeifle} {et~al.}(2022){Pfeifle}, {Satyapal}, {Ricci}, {Secrest},
  {Gliozzi}, {Bohn}, {Canalizo}, \& {Reefe}}]{pfeifle2022b}
{Pfeifle}, R.~W., {Satyapal}, S., {Ricci}, C., {et~al.} 2022, arXiv e-prints,
  arXiv:2211.17271.
\newblock \doarXiv{2211.17271}

\bibitem[{{Pfeifle} {et~al.}(2019{\natexlab{a}}){Pfeifle}, {Satyapal},
  {Secrest}, {Gliozzi}, {Ricci}, {Ellison}, {Rothberg}, {Cann}, {Blecha},
  {Williams}, \& {Constantin}}]{pfeifle2019a}
{Pfeifle}, R.~W., {Satyapal}, S., {Secrest}, N.~J., {et~al.}
  2019{\natexlab{a}}, \apj, 875, 117, \dodoi{10.3847/1538-4357/ab07bc}

\bibitem[{{Pfeifle} {et~al.}(2019{\natexlab{b}}){Pfeifle}, {Satyapal},
  {Manzano-King}, {Cann}, {Sexton}, {Rothberg}, {Canalizo}, {Ricci}, {Blecha},
  {Ellison}, {Gliozzi}, {Secrest}, {Constantin}, \& {Harvey}}]{pfeifle2019b}
{Pfeifle}, R.~W., {Satyapal}, S., {Manzano-King}, C., {et~al.}
  2019{\natexlab{b}}, \apj, 883, 167, \dodoi{10.3847/1538-4357/ab3a9b}

\bibitem[{{Pfeifle} {et~al.}(2021){Pfeifle}, {Ricci}, {Boorman}, {Stalevski},
  {Asmus}, {Trakhtenbrot}, {Koss}, {Stern}, {Ricci}, {Satyapal}, {Ichikawa},
  {Rosario}, {Caglar}, {Treister}, {Powell}, {Oh}, {Urry}, \&
  {Harrison}}]{pfeifle2022}
{Pfeifle}, R.~W., {Ricci}, C., {Boorman}, P.~G., {et~al.} 2021, arXiv e-prints,
  arXiv:2102.04412.
\newblock \doarXiv{2102.04412}

\bibitem[{{Piconcelli} {et~al.}(2010){Piconcelli}, {Vignali}, {Bianchi},
  {Mathur}, {Fiore}, {Guainazzi}, {Lanzuisi}, {Maiolino}, \&
  {Nicastro}}]{piconcelli2010}
{Piconcelli}, E., {Vignali}, C., {Bianchi}, S., {et~al.} 2010, \apjl, 722,
  L147, \dodoi{10.1088/2041-8205/722/2/L147}

\bibitem[{{Powell} {et~al.}(2018){Powell}, {Cappelluti}, {Urry}, {Koss},
  {Finoguenov}, {Ricci}, {Trakhtenbrot}, {Allevato}, {Ajello}, {Oh},
  {Schawinski}, \& {Secrest}}]{2018ApJ...858..110P}
{Powell}, M.~C., {Cappelluti}, N., {Urry}, C.~M., {et~al.} 2018, \apj, 858,
  110, \dodoi{10.3847/1538-4357/aabd7f}

\bibitem[{{Ptak} {et~al.}(2015){Ptak}, {Hornschemeier}, {Zezas}, {Lehmer},
  {Yukita}, {Wik}, {Antoniou}, {Argo}, {Ballo}, {Bechtol}, {Boggs}, {Della
  Ceca}, {Christensen}, {Craig}, {Hailey}, {Harrison}, {Krivonos}, {Maccarone},
  {Stern}, {Tatum}, {Venters}, \& {Zhang}}]{ptak2015}
{Ptak}, A., {Hornschemeier}, A., {Zezas}, A., {et~al.} 2015, \apj, 800, 104,
  \dodoi{10.1088/0004-637X/800/2/104}

\bibitem[{{Ranalli} {et~al.}(2003){Ranalli}, {Comastri}, \&
  {Setti}}]{ranalli2003}
{Ranalli}, P., {Comastri}, A., \& {Setti}, G. 2003, \aap, 399, 39,
  \dodoi{10.1051/0004-6361:20021600}

\bibitem[{{Ricci} {et~al.}(2015){Ricci}, {Ueda}, {Koss}, {Trakhtenbrot},
  {Bauer}, \& {Gandhi}}]{ricci2015}
{Ricci}, C., {Ueda}, Y., {Koss}, M.~J., {et~al.} 2015, \apjl, 815, L13,
  \dodoi{10.1088/2041-8205/815/1/L13}

\bibitem[{{Ricci} {et~al.}(2017{\natexlab{a}}){Ricci}, {Bauer}, {Treister},
  {Schawinski}, {Privon}, {Blecha}, {Arevalo}, {Armus}, {Harrison}, {Ho},
  {Iwasawa}, {Sanders}, \& {Stern}}]{ricci2017MNRAS}
{Ricci}, C., {Bauer}, F.~E., {Treister}, E., {et~al.} 2017{\natexlab{a}},
  \mnras, 468, 1273, \dodoi{10.1093/mnras/stx173}

\bibitem[{{Ricci} {et~al.}(2017{\natexlab{b}}){Ricci}, {Trakhtenbrot}, {Koss},
  {Ueda}, {Del Vecchio}, {Treister}, {Schawinski}, {Paltani}, {Oh}, {Lamperti},
  {Berney}, {Gandhi}, {Ichikawa}, {Bauer}, {Ho}, {Asmus}, {Beckmann}, {Soldi},
  {Balokovi{\'c}}, {Gehrels}, \& {Markwardt}}]{ricci2017bass}
{Ricci}, C., {Trakhtenbrot}, B., {Koss}, M.~J., {et~al.} 2017{\natexlab{b}},
  \apjs, 233, 17, \dodoi{10.3847/1538-4365/aa96ad}

\bibitem[{{Ricci} {et~al.}(2021){Ricci}, {Privon}, {Pfeifle}, {Armus},
  {Iwasawa}, {Torres-Alb{\`a}}, {Satyapal}, {Bauer}, {Treister}, {Ho}, {Aalto},
  {Ar{\'e}valo}, {Barcos-Mu{\~n}oz}, {Charmandaris}, {Diaz-Santos}, {Evans},
  {Gao}, {Inami}, {Koss}, {Lansbury}, {Linden}, {Medling}, {Sanders}, {Song},
  {Stern}, {U}, {Ueda}, \& {Yamada}}]{ricci2021}
{Ricci}, C., {Privon}, G.~C., {Pfeifle}, R.~W., {et~al.} 2021, \mnras, 506,
  5935, \dodoi{10.1093/mnras/stab2052}

\bibitem[{{Robitaille} \& {Bressert}(2012)}]{robitaille2012}
{Robitaille}, T., \& {Bressert}, E. 2012, {APLpy: Astronomical Plotting Library
  in Python}, Astrophysics Source Code Library, record ascl:1208.017.
\newblock \doeprint{1208.017}

\bibitem[{{Satyapal} {et~al.}(2014){Satyapal}, {Secrest}, {McAlpine},
  {Ellison}, {Fischer}, \& {Rosenberg}}]{satyapal2014}
{Satyapal}, S., {Secrest}, N.~J., {McAlpine}, W., {et~al.} 2014, \apj, 784,
  113, \dodoi{10.1088/0004-637X/784/2/113}

\bibitem[{{Satyapal} {et~al.}(2017){Satyapal}, {Secrest}, {Ricci}, {Ellison},
  {Rothberg}, {Blecha}, {Constantin}, {Gliozzi}, {McNulty}, \&
  {Ferguson}}]{satyapal2017}
{Satyapal}, S., {Secrest}, N.~J., {Ricci}, C., {et~al.} 2017, \apj, 848, 126,
  \dodoi{10.3847/1538-4357/aa88ca}

\bibitem[{{Schlafly} \& {Finkbeiner}(2011)}]{2011ApJ...737..103S}
{Schlafly}, E.~F., \& {Finkbeiner}, D.~P. 2011, \apj, 737, 103,
  \dodoi{10.1088/0004-637X/737/2/103}

\bibitem[{{Secrest} {et~al.}(2017){Secrest}, {Schmitt}, {Blecha}, {Rothberg},
  \& {Fischer}}]{secrest2017}
{Secrest}, N.~J., {Schmitt}, H.~R., {Blecha}, L., {Rothberg}, B., \& {Fischer},
  J. 2017, \apj, 836, 183, \dodoi{10.3847/1538-4357/836/2/183}

\bibitem[{{Skrutskie} {et~al.}(2006){Skrutskie}, {Cutri}, {Stiening},
  {Weinberg}, {Schneider}, {Carpenter}, {Beichman}, {Capps}, {Chester},
  {Elias}, {Huchra}, {Liebert}, {Lonsdale}, {Monet}, {Price}, {Seitzer},
  {Jarrett}, {Kirkpatrick}, {Gizis}, {Howard}, {Evans}, {Fowler}, {Fullmer},
  {Hurt}, {Light}, {Kopan}, {Marsh}, {McCallon}, {Tam}, {Van Dyk}, \&
  {Wheelock}}]{2006AJ....131.1163S}
{Skrutskie}, M.~F., {Cutri}, R.~M., {Stiening}, R., {et~al.} 2006, \aj, 131,
  1163, \dodoi{10.1086/498708}

\bibitem[{{Smith} {et~al.}(2010){Smith}, {Shields}, {Bonning}, {McMullen},
  {Rosario}, \& {Salviander}}]{smith2010}
{Smith}, K.~L., {Shields}, G.~A., {Bonning}, E.~W., {et~al.} 2010, \apj, 716,
  866, \dodoi{10.1088/0004-637X/716/1/866}

\bibitem[{{Stern}(2015)}]{stern2015}
{Stern}, D. 2015, \apj, 807, 129, \dodoi{10.1088/0004-637X/807/2/129}

\bibitem[{{Stern} {et~al.}(2012){Stern}, {Assef}, {Benford}, {Blain}, {Cutri},
  {Dey}, {Eisenhardt}, {Griffith}, {Jarrett}, {Lake}, {Masci}, {Petty},
  {Stanford}, {Tsai}, {Wright}, {Yan}, {Harrison}, \& {Madsen}}]{stern2012}
{Stern}, D., {Assef}, R.~J., {Benford}, D.~J., {et~al.} 2012, \apj, 753, 30,
  \dodoi{10.1088/0004-637X/753/1/30}

\bibitem[{{Taylor}(2005)}]{2005ASPC..347...29T}
{Taylor}, M.~B. 2005, in Astronomical Society of the Pacific Conference Series,
  Vol. 347, Astronomical Data Analysis Software and Systems XIV, ed.
  P.~{Shopbell}, M.~{Britton}, \& R.~{Ebert}, 29

\bibitem[{{Torres-Alb{\`a}} {et~al.}(2018){Torres-Alb{\`a}}, {Iwasawa},
  {D{\'\i}az-Santos}, {Charmandaris}, {Ricci}, {Chu}, {Sanders}, {Armus},
  {Barcos-Mu{\~n}oz}, {Evans}, {Howell}, {Inami}, {Linden}, {Medling},
  {Privon}, {U}, \& {Yoon}}]{torres-alba2018}
{Torres-Alb{\`a}}, N., {Iwasawa}, K., {D{\'\i}az-Santos}, T., {et~al.} 2018,
  \aap, 620, A140, \dodoi{10.1051/0004-6361/201834105}

\bibitem[{{Tozzi} {et~al.}(2006){Tozzi}, {Gilli}, {Mainieri}, {Norman},
  {Risaliti}, {Rosati}, {Bergeron}, {Borgani}, {Giacconi}, {Hasinger},
  {Nonino}, {Streblyanska}, {Szokoly}, {Wang}, \& {Zheng}}]{tozzi2006}
{Tozzi}, P., {Gilli}, R., {Mainieri}, V., {et~al.} 2006, \aap, 451, 457,
  \dodoi{10.1051/0004-6361:20042592}

\bibitem[{van~der Walt {et~al.}(2011)van~der Walt, Colbert, \&
  Varoquaux}]{walt2011}
van~der Walt, S., Colbert, S.~C., \& Varoquaux, G. 2011, Computing in Science
  and Engineering, 13, 22.
\newblock \url{http://dblp.uni-trier.de/db/journals/cse/cse13.html#WaltCV11}

\bibitem[{{Van Wassenhove} {et~al.}(2012){Van Wassenhove}, {Volonteri},
  {Mayer}, {Dotti}, {Bellovary}, \& {Callegari}}]{vanwassenhove2012}
{Van Wassenhove}, S., {Volonteri}, M., {Mayer}, L., {et~al.} 2012, \apjl, 748,
  L7, \dodoi{10.1088/2041-8205/748/1/L7}

\bibitem[{{Virtanen} {et~al.}(2020){Virtanen}, {Gommers}, {Oliphant},
  {Haberland}, {Reddy}, {Cournapeau}, {Burovski}, {Peterson}, {Weckesser},
  {Bright}, {van der Walt}, {Brett}, {Wilson}, {Jarrod Millman}, {Mayorov},
  {Nelson}, {Jones}, {Kern}, {Larson}, {Carey}, {Polat}, {Feng}, {Moore}, {Vand
  erPlas}, {Laxalde}, {Perktold}, {Cimrman}, {Henriksen}, {Quintero}, {Harris},
  {Archibald}, {Ribeiro}, {Pedregosa}, {van Mulbregt}, \&
  {Contributors}}]{virtanen2020}
{Virtanen}, P., {Gommers}, R., {Oliphant}, T.~E., {et~al.} 2020, Nature
  Methods, \dodoi{https://doi.org/10.1038/s41592-019-0686-2}

\bibitem[{{Wang} {et~al.}(2009){Wang}, {Chen}, {Hu}, {Mao}, {Zhang}, \&
  {Bian}}]{wang2009}
{Wang}, J.-M., {Chen}, Y.-M., {Hu}, C., {et~al.} 2009, \apjl, 705, L76,
  \dodoi{10.1088/0004-637X/705/1/L76}

\bibitem[{{Willingale} {et~al.}(2013){Willingale}, {Starling}, {Beardmore},
  {Tanvir}, \& {O'Brien}}]{willingale2013}
{Willingale}, R., {Starling}, R.~L.~C., {Beardmore}, A.~P., {Tanvir}, N.~R., \&
  {O'Brien}, P.~T. 2013, \mnras, 431, 394, \dodoi{10.1093/mnras/stt175}

\bibitem[{{Wright} {et~al.}(2010){Wright}, {Eisenhardt}, {Mainzer}, {Ressler},
  {Cutri}, {Jarrett}, {Kirkpatrick}, {Padgett}, {McMillan}, {Skrutskie},
  {Stanford}, {Cohen}, {Walker}, {Mather}, {Leisawitz}, {Gautier}, {McLean},
  {Benford}, {Lonsdale}, {Blain}, {Mendez}, {Irace}, {Duval}, {Liu}, {Royer},
  {Heinrichsen}, {Howard}, {Shannon}, {Kendall}, {Walsh}, {Larsen}, {Cardon},
  {Schick}, {Schwalm}, {Abid}, {Fabinsky}, {Naes}, \&
  {Tsai}}]{2010AJ....140.1868W}
{Wright}, E.~L., {Eisenhardt}, P. R.~M., {Mainzer}, A.~K., {et~al.} 2010, \aj,
  140, 1868, \dodoi{10.1088/0004-6256/140/6/1868}

\bibitem[{{Wright} {et~al.}(2019){Wright}, {Eisenhardt}, {Mainzer}, {Ressler},
  {Cutri}, {Jarrett}, {Kirkpatrick}, {Padgett}, {McMillan}, {Skrutskie},
  {Stanford}, {Cohen}, {Walker}, {Mather}, {Leisawitz}, {Gautier}, {McLean},
  {Benford}, {Lonsdale}, {Blain}, {Mendez}, {Irace}, {Duval}, {Liu}, {Royer},
  {Heinrichsen}, {Howard}, {Shannon}, {Kendall}, {Walsh}, {Larsen}, {Cardon},
  {Schick}, {Schwalm}, {Abid}, {Fabinsky}, {Naes}, \&
  {Tsai}}]{https://doi.org/10.26131/irsa1}
---. 2019, AllWISE Source Catalog,  IPAC, \dodoi{10.26131/IRSA1}

\bibitem[{{Yamada} {et~al.}(2018){Yamada}, {Ueda}, {Oda}, {Tanimoto},
  {Imanishi}, {Terashima}, \& {Ricci}}]{yamada2018}
{Yamada}, S., {Ueda}, Y., {Oda}, S., {et~al.} 2018, \apj, 858, 106,
  \dodoi{10.3847/1538-4357/aabacb}

\bibitem[{{Yamada} {et~al.}(2021){Yamada}, {Ueda}, {Tanimoto}, {Imanishi},
  {Toba}, {Ricci}, \& {Privon}}]{yamada2021}
{Yamada}, S., {Ueda}, Y., {Tanimoto}, A., {et~al.} 2021, \apjs, 257, 61,
  \dodoi{10.3847/1538-4365/ac17f5}

\end{thebibliography}
\bibliographystyle{aasjournal}

\end{document}